%% file: secdec3_main.tex
\documentclass[11pt,1p]{myelsart}
\usepackage{graphics,color,epsfig,amsmath,amssymb}
\usepackage{multirow,bigdelim,lscape}
\usepackage{feynmp}
\usepackage{subfigure}
\usepackage{xspace}
\usepackage{listings}
\usepackage{cite}
\graphicspath{{figures/}}
\newcounter{bla}

\def\be{\begin{align}}
\def\ee{\end{align}}
\def\bea{\begin{align}}
\def\eea{\end{align}}
\def\nn{\nonumber}

\newcommand{\secdec}{\textsc{SecDec}\xspace}
\newcommand{\idel}{i\,\delta}
\newcommand{\eps}{\epsilon}
\newcommand{\rd}{{\mathrm{d}}}

\newcommand{\dd}[1]{\mathrm{d}#1\,}
\newcommand{\deltafun}{\,\delta}

\begin{document}

\begin{frontmatter}
\hfill{\scriptsize{FR-PHENO-2015-001, MPP-2015-27, ZU-TH 1/15}}\\

\title{SecDec-3.0: numerical evaluation of multi-scale integrals beyond one loop}

\author[a]{S.~Borowka},
\author[b]{G.~Heinrich},
\author[b,c]{S.~P.~Jones},
\author[b]{M.~Kerner},
\author[b]{J.~Schlenk},
\author[b]{T.~Zirke}

\address[a]{Institute for Physics, Universit{\"a}t Z{\"u}rich, Winterthurerstr.190, 8057 Z\"urich,
Switzerland}
\address[b]{Max-Planck-Institute for Physics, F\"ohringer Ring 6, 80805 M\"unchen, Germany}
\address[c]{Albert-Ludwigs-Universit\"at Freiburg, Physikalisches Institut, 79104
Freiburg, Germany}

\begin{abstract}
\secdec{} is a program which can be used for the 
factorization of dimensionally regulated poles from parametric integrals, 
in particular multi-loop integrals,
and the subsequent numerical evaluation of the finite coefficients.
Here we present version 3.0 of the program, which has major improvements 
compared to version 2: it is faster, contains new decomposition strategies, 
an improved user interface and various other new features
which extend the range of applicability.
\end{abstract}



\end{frontmatter}

\newpage

{\bf PROGRAM SUMMARY}

\begin{small}
\noindent
{\em Program Title: } SecDec 3.0                                          \\
{\em Journal Reference:}  Comput. Phys. Comm. 196 (2015) 470.                               \\
{\em Catalogue identifier:}     AEIR\_v3\_0                               \\
{\em Licensing provisions:} Standard CPC license                                 \\
{\em Programming language:} Wolfram Mathematica, perl, Fortran/C++       \\
{\em Computer:}  from a single PC to a cluster, depending on the problem  \\
{\em Operating system: } Unix, Linux                                       \\
{\em RAM:} depending on the complexity of the problem                                              
 \\
{\em Classification:}                                         \\
  4.4 Feynman diagrams, 
  5 Computer Algebra, 
  11.1 General, High Energy Physics and Computing.\\
{\em Journal reference of previous version:} Comput. Phys. Commun. 184 (2013) 2552. \\
{\em Nature of the problem:}\\
  Extraction of ultraviolet and infrared singularities from parametric integrals 
  appearing in higher order perturbative calculations in gauge theories. 
  Numerical integration in the presence of integrable singularities 
  (e.g. kinematic thresholds). \\
{\em Solution method:}\\
 Algebraic extraction of singularities within dimensional regularization using iterated sector decomposition. 
 This leads to a Laurent series in the dimensional regularization parameter $\epsilon$, 
 where the coefficients are finite integrals over the unit-hypercube. 
 Those integrals are evaluated numerically by Monte Carlo integration.
 The integrable singularities are handled by 
 choosing a suitable integration contour in the complex plane, in an automated way.
   \\
{\em Restrictions:} Depending on the complexity of the problem, limited by 
memory and CPU time.\\
{\em Running time:}\\
Between a few seconds and several hours, depending on the complexity of the problem.\\
\end{small}

\newpage


\section{Introduction}
\label{sec:intro}
\input{intro}
%
\section{Theoretical framework}
\label{sec:theo}
\input{theo}
\section{Structure and new features  of \secdec version 3.0}
\label{sec:program}
\input{program}

\section{Installation and usage}
\label{sec:instusage}
\input{usage}
%
\section{Examples}
\label{sec:examples}
\input{examples}

\vspace*{3mm}

\section{Conclusions}
\label{sec:conclusion}
We have presented version 3.0 of the program {\sc SecDec}, 
which is publicly available at 
{\tt http://secdec.hepforge.org}.\\
The part of the program which allows to calculate multi-loop integrals 
for arbitrary kinematics has been improved in various respects:
it contains two additional 
decomposition strategies which are guaranteed to stop, based on a geometric algorithm.
In addition, it can deal with tensor integrals in the form of inverse propagators, 
or, more generally, can take lists of indices where the indices can also be 
negative or zero. 
Integrals containing linear propagators can also be calculated.
The timings also have been improved, and the possibility to do the numerical integrations 
for a large range of kinematic points on a cluster has been made much more 
straightforward. 
Further, it is now possible to use an integrator from Mathematica in addition to the {\sc Cuba} library. 
Together with the re-structuring of the code, 
the whole interface has been made more user-friendly.\\
The part of the program which allows to factorize poles from parameter integrals which are not related to
loop integrals also contains a new feature,  allowing to introduce some dummy functions which 
will not be decomposed, but can themselves depend on $\eps$. 

With all its new features and the new user interface, we believe that {\sc SecDec}-3.0  
will certainly be useful for a number of calculations beyond NLO, in particular 
in cases where several mass scales are involved, and where analytical calculations 
of the loop integrals are at their limit.

\section*{Acknowledgements}
We would like to thank Thomas Hahn and Johann Felix von Soden-Fraunhofen for helpful
discussions and suggestions, and Anton Stoyanov for useful tests.
We also would like to thank Christoph Greub, Peter Uwer and Dominik Kara for valuable comments.
This reserch was supported in part by the 
Research Executive Agency (REA) of the European Union under the Grant Agreement
PITN-GA2012316704 (HiggsTools).
S. Borowka gratefully acknowledges financial support by the ERC Advanced Grant MC@NNLO (340983). 

\renewcommand \thesection{\Alph{section}}
\renewcommand{\theequation}{\Alph{section}.\arabic{equation}}
\setcounter{section}{0}
\setcounter{equation}{0}

\section{User Manual}
\label{sec:appendix:manual}

\input{appendix}



\input secdec3_main.bbl

\end{document}

%% file: intro.tex
After the very successful Run I of the LHC, pushing the precision frontier is and will be one of the primary goals
for the next phase of LHC data taking and at future colliders.

This means that higher-order corrections within the Standard Model, and promising extensions beyond, 
need to be evaluated. In addition, in order to scrutinize the Higgs properties, 
in particular the Yukawa couplings to fermions, 
heavy quark masses should be taken into account without resorting to low energy approximations.
Therefore, it is of primary importance to have tools at hand for the calculation of two- (and more) loop
integrals involving several mass scales. 

In  general, 
high precision calculations have in common 
that they involve multi-dimensional integrations over some parameters: 
Feynman (or Schwinger) parameters in the case of (multi-)loop integrals, 
or parameters related to the integration 
of subtraction terms over a factorised phase space in the case of infrared-divergent 
real radiation.
Usually, these calculations are performed within the framework of dimensional regularization, 
and one of the challenges is to factorise the poles in the regulator $\eps$. 

The program \secdec\,\cite{Carter:2010hi,Borowka:2012yc,Borowka:2013cma} is designed to 
perform this task in an automated way, and to integrate the 
coefficients of the resulting Laurent series in $\eps$ numerically,
based on the sector decomposition algorithm described in Refs.~\cite{Binoth:2000ps,Heinrich:2008si},
which was inspired by earlier ideas as contained in Refs.~\cite{Hepp:1966eg,Roth:1996pd}.
Other public implementations of sector decomposition can be 
found in Refs.~\cite{Bogner:2007cr,Gluza:2010rn,Smirnov:2008py,Smirnov:2009pb,Smirnov:2013eza}. 

The numerical integration in {\sc SecDec}-1.0~\cite{Carter:2010hi} was restricted to Euclidean kinematics 
for integrals with more than one kinematic scale. However, this restriction was lifted 
in {\sc SecDec}-2.0~\cite{Borowka:2012yc,Borowka:2014aaa}, by combining sector decomposition with a method to deform 
the multi-dimensional integration contour into the complex plane~\cite{Soper:1999xk}.
While such a method of contour deformation already had been applied in various contexts --
for examples at one loop see e.g. Refs.~\cite{Binoth:2002xh,Nagy:2006xy,Binoth:2005ff,Lazopoulos:2007ix,Lazopoulos:2007bv,Gong:2008ww,Becker:2010ng}, 
for two-loop examples  Refs.~\cite{Anastasiou:2006hc,Anastasiou:2007qb,Beerli:2008zz} --
it has been combined with the automated setup for the resolution of singularities and 
made publicly available for the first time in Ref.~\cite{Borowka:2012yc}. 

Very recently, another strategy to achieve a resolution of dimensionally 
regulated  singularities in multi-loop integrals has been proposed~\cite{Panzer:2014gra,vonManteuffel:2014qoa},
which utilizes integration by parts and dimension shifts.

The original sector decomposition algorithm described in Ref.~\cite{Binoth:2000ps} is based on 
an iterative procedure, which can run into an infinite recursion in certain (rather rare) 
situations. It was soon noticed~\cite{Bogner:2007cr} however that the structure of Feynman
integrals is such that a decomposition algorithm must exist which is guaranteed to stop, 
as the procedure can be mapped to a well known problem in convex geometry.
Related observations have been made in Ref.~\cite{Smirnov:2008aw}. 
In Ref.~\cite{Kaneko:2009qx}, an algorithm was presented which cannot lead to infinite recursion 
and is more efficient than previous algorithms with this property. 

In this paper, we present the implementation of two new decomposition strategies as alternatives to 
the already implemented heuristic algorithm. They are based on the method of 
Ref.~\cite{Kaneko:2009qx} and therefore guaranteed to avoid infinite recursion.
This is described in Section~\ref{sec:program}, together with other new features of the program. 
As the user interface has been restructured in version 3 of \secdec, 
we describe its usage in detail in Section~\ref{sec:instusage} and give various examples in Section~\ref{sec:examples}. 
After the conclusions, an appendix provides a detailed description of the various input options and run modes.

%% file: theo.tex
In this section we give the expressions for the representation of multi-loop integrals in
terms of Feynman parameters and explain our notation. 

\subsection{Feynman integrals}
In order to define our conventions we choose a scalar integral. 
Integrals with loop momenta in the numerator
only lead to an additional function of the Feynman parameters and invariants 
in the numerator and will be discussed in Section~\ref{sec:inversepropagators}.

A scalar Feynman integral $G$ in $D$ dimensions 
at $L$ loops with  $N$ propagators, where 
the propagators can have arbitrary, not necessarily integer powers $\nu_j$,  
has the following representation in momentum space:
\begin{eqnarray}\label{eq:integraldef}
G&=&\int\prod\limits_{l=1}^{L} \rd^D\kappa_l\;
\frac{1}
{\prod\limits_{j=1}^{N} P_{j}^{\nu_j}(\{k\},\{p\},m_j^2)}\\
\rd^D\kappa_l&=&\frac{\mu^{4-D}}{i\pi^{\frac{D}{2}}}\,\rd^D k_l\;,\;
P_j(\{k\},\{p\},m_j^2)=q_j^2-m_j^2+i\delta\;,\nn
\end{eqnarray}
where the $q_j$ are linear combinations of external momenta $p_i$ and loop momenta $k_l$.
The numerical result produced by \secdec will correspond to $G$ with $\mu=1$, 
except if the user specifies a different prefactor when defining the graph to be calculated.

Denominator factors which only have a linear, but not a quadratic dependence on the loop momenta 
can appear in various contexts within loop calculations. In the new version presented here, \secdec 
can deal with such propagators. However, care has to be taken with the analytic continuation 
in this case, as will be discussed in Section~\ref{subsec:linprops}.

Introducing Feynman parameters in Eq.~(\ref{eq:integraldef}) leads to
\begin{eqnarray}
G&=&  
\frac{\Gamma(N_\nu)}{\prod_{j=1}^{N}\Gamma(\nu_j)}
\int_0^\infty \,\prod\limits_{j=1}^{N}\rd x_j\,\,x_j^{\nu_j-1}\, 
\delta\big(1-\sum_{i=1}^N x_i\big)\nn \\
&&\cdot \int \rd^D\kappa_1\ldots\rd^D\kappa_L
\left[ 
       \sum\limits_{i,j=1}^{L} k_i^{\rm{T}}\, M_{ij}\, k_j  - 
       2\sum\limits_{j=1}^{L} k_j^{\rm{T}}\cdot Q_j +J +\idel
                             \right]^{-N\nu}\\
&&\nn\\
&=&\frac{(-1)^{N_{\nu}}}{\prod_{j=1}^{N}\Gamma(\nu_j)}\Gamma(N_{\nu}-LD/2)\nn\\
&&\cdot \int\limits_{0}^{\infty} 
\,\prod\limits_{j=1}^{N}\text{d}x_j\,\,x_j^{\nu_j-1}\,\delta(1-\sum_{l=1}^N x_l)
\frac{{\cal U}^{N_{\nu}-(L+1) D/2}}
{{\cal F}^{N_\nu-L D/2}}\;,\label{eq:scalarloopint}\\
 &&\nonumber\\
&&\mbox{where}  \nonumber\\
{\cal F}(\vec x) &=& \det (M) 
\left[ \sum\limits_{j,l=1}^{L} Q_j \, M^{-1}_{jl}\, Q_l
-J -\idel\right]\label{DEF:F}\\
{\cal U}(\vec x) &=& \det (M),  \; N_\nu=\sum_{j=1}^N\nu_j\;.\label{DEF:U}
\end{eqnarray}
In the expressions above, $M$ is an $L\times L$ matrix containing Feynman parameters,
$Q$ is an $L$-dimensional vector, where each entry is a linear combination of external momenta and 
Feynman parameters, and $J$ is a scalar expression containing kinematic invariants and Feynman parameters.
For more details and examples, see e.g. Ref.~\cite{Binoth:2000ps}.


The functions ${\cal U}$ and ${\cal F}$, called graph polynomials in the
following, can also be constructed
from the topology of the corresponding 
Feynman graph~\cite{Tarasov:1996br,Smirnov:2006ry,Heinrich:2008si}. 
In \secdec, this is implemented as one way to find the two graph polynomials. 

For a diagram with massless propagators, 
none of the Feynman parameters occurs quadratically in 
the function ${\cal F}={\cal F}_0$. If massive internal lines are present, 
${\cal F}$ gets an additional term 
${\cal F}(\vec x) =  {\cal F}_0(\vec x) + {\cal U}(\vec x) \sum\limits_{j=1}^{N} x_j m_j^2$.

${\cal U}$ is a positive semi-definite function. 
A vanishing ${\cal U}$ function is related to the UV subdivergences of the graph, 
a vanishing ${\cal F}$ function to its IR divergences. 
In the region where all invariants formed from external momenta are negative, 
which we will call the {\em Euclidean region} in the following, 
${\cal F}$ is also a positive semi-definite function 
of the Feynman parameters $x_j$.  If some of the invariants are zero, 
for example if some of the external momenta
are light-like, an IR divergence may appear due to ${\cal F}$ vanishing.
Therefore it depends on the {\em kinematics},
and not only on the topology (like in the UV case), 
whether a zero of ${\cal F}$ leads to a divergence or not. 
The necessary (but not sufficient) conditions for a divergence 
are given by the Landau equations~\cite{Landau:1959fi,Nakanishi}.
If all kinematic invariants formed by external momenta are negative, 
the necessary condition ${\cal F}=0$ for an IR divergence can only 
be fulfilled if some of the parameters $x_i$ are zero.
These endpoint singularities can be regulated by dimensional regularisation and 
factored out of the function ${\cal F}$ using sector decomposition. 
The same holds for dimensionally regulated UV singularities contained in ${\cal U}$.
However, after the UV and IR  singularities have been extracted 
as poles in  1/$\epsilon$,
for non-Euclidean kinematics 
integrable singularities related to kinematic thresholds remain. 
These singularities imply that ${\cal F}$ is vanishing inside the integration region 
for some combinations of Feynman parameter values and values of the kinematic invariants.
However, the integrals can be evaluated 
by deforming the integration contour into the
complex plane, as explained in detail in Ref.~\cite{Borowka:2012yc}.

The numerical result given by \secdec will be the one for the integral as defined in 
Eq.~(\ref{eq:integraldef}).
This implies that the prefactor $\frac{(-1)^{N_{\nu}}}{\prod_{j=1}^{N}\Gamma(\nu_j)}\Gamma(N_{\nu}-LD/2)$ 
coming from the Feynman parametrisation  
by default will be {\it included} in our numerical result. 
However, the user can define a different prefactor to be factored out from the numerical
result. 

%% file: program.tex
\subsection{Program structure}

The program consists of two main parts, one designed for loop integrals, 
the other for more general parametric integrals. 
The sector decomposition procedure to factorize the poles in the regulator $\eps$, 
and the way the subtractions are done, is common to both parts.
However, contour deformation is only supported for loop integrals, as the analytic continuation 
can be performed in an automated way, following Feynman's $i \delta$ prescription. 
More general, user-defined loop integral functions which share the analytic continuation 
properties of loop integrals can also be treated 
in the {\tt loop} directory, see Section~\ref{subsec:exampleuserdefined} for a detailed example.

\begin{figure}[htb]
\begin{center}
\includegraphics[width=0.85\textwidth]{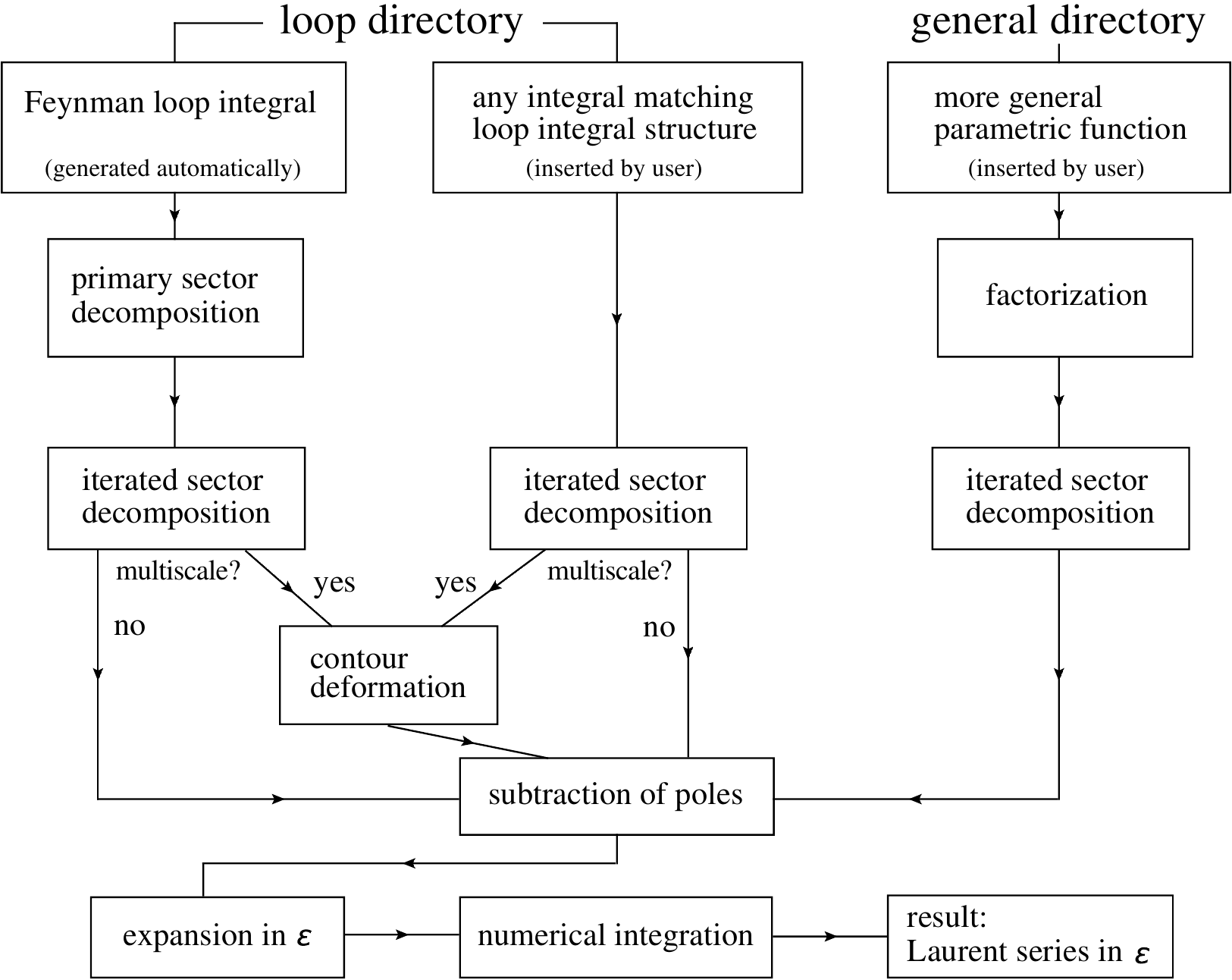}
\end{center}
\caption{Flowchart showing the main steps the program performs to produce the 
	numerical result as a Laurent series in $\epsilon$.
	\label{fig:structure} }
\end{figure}

\begin{figure}[htb]
\begin{picture}(300,200)(0,0)
\put(-75,-480){
\includegraphics[width=20cm]{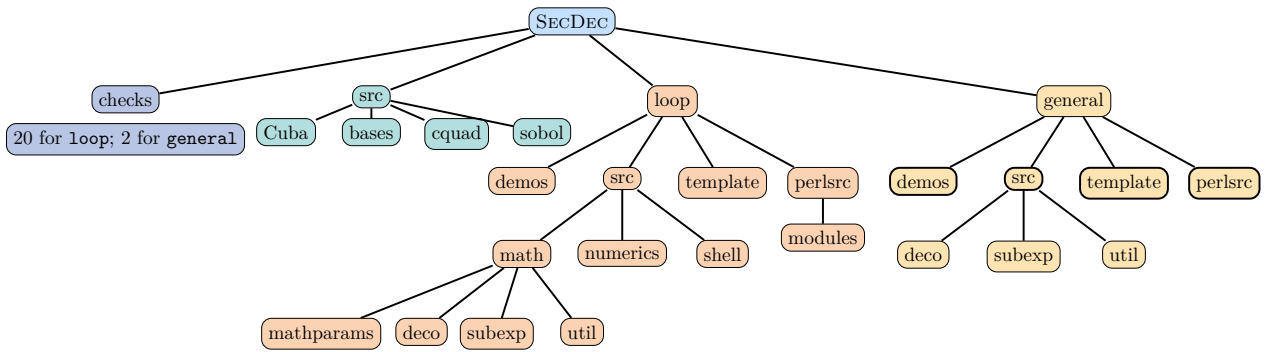}}
\end{picture}
\caption{Directory structure of \secdec version 3.
	\label{fig:dirstructure} }
\end{figure}

The basic operational sequence of the program is shown in Fig.~\ref{fig:structure}. 
We also show a diagram for the internal \secdec directory structure in Fig.~\ref{fig:dirstructure}.

\clearpage
\subsection{New features}
\subsubsection{New user interface}
\label{sec:newinterface}

The very first change to mention is that the command to launch \secdec 
is now {\tt secdec} rather than {\tt launch}.

The restructuring of the user interface described below mainly applies to the {\tt loop} 
setup. In the {\tt general} branch(=folder), the program is now also 
launched by the command {\tt secdec}, but otherwise the user interface is still the same 
as in version 2. 
The user interface in the {\tt loop} branch 
has been restructured with the following aims in mind:
\begin{itemize}
\item facilitating scans over large ranges of kinematic parameters
\item making the usage of \secdec on a cluster straightforward
\item facilitating interfaces to reduction codes like {\sc
    Reduze}~\cite{vonManteuffel:2012np},  {\sc
    Fire}~\cite{Smirnov:2014hma} or {\sc LiteRed}~\cite{Lee:2012cn}
\item allowing the user to define his/her own names for kinematic invariants
\item reducing the mandatory user input to a minimum.
\end{itemize}
For these reasons, we have decoupled the definition of the numerical values for the 
kinematic invariants from the file {\tt paramloop.input} and introduced a separate 
input file {\tt kinem.input} to define the numerical values for the kinematic points to be calculated. 
Each line in {\tt kinem.input} denotes a new kinematic point. 
The usage is described in detail in Section \ref{sec:usage} and in Appendix \ref{appendix:loop}.

We also changed parts of the Mathematica file which serves to define the graph to be calculated. 
The default name has changed from {\tt templateloop.m} to {\tt math.m} 
(however, the user can still give it any name, e.g. {\tt mygraph.m}).
The default name for the parameter input file is now {\tt param.input}, 
the default name for the file defining the numerical values for the kinematic invariants is {\tt kinem.input}.
Accordingly, the command to call \secdec is now either\\
{\tt {\bf secdec} }\\
if the default names for the input files are used, or\\
{\tt {\bf secdec} -p <myparam.input>  -m <mygraph.m> -k <mykinem.input>}.\\
Please note that what has been the ``{\tt -t}" option is now renamed into the ``{\tt -m}" option.

The main changes in the graph definition file {\tt math.m} are 
\begin{itemize}
\item The list of propagator powers,  {\tt powerlist}, (also called ``indices" in the literature)
can now also contain zero or negative entries, where the latter correspond to inverse propagators forming a numerator.
Such integrals are calculated using the algorithm described in Section~\ref{sec:inversepropagators}.
This way the interface to reduction programs providing the master integrals in the form of 
lists of indices is straightforward.
\item The symbols for the kinematic invariants are not predefined, but can be defined by the user.
This also means that the user needs to define the expressions for the scalar products of 
Lorentz vectors (in {\tt ScalarProductRules}) ocurring in the graph.
\end{itemize}

\begin{figure}[htb]
\begin{picture}(300,400)(0,0)
\put(-65,-300){\includegraphics[width=19.5cm]{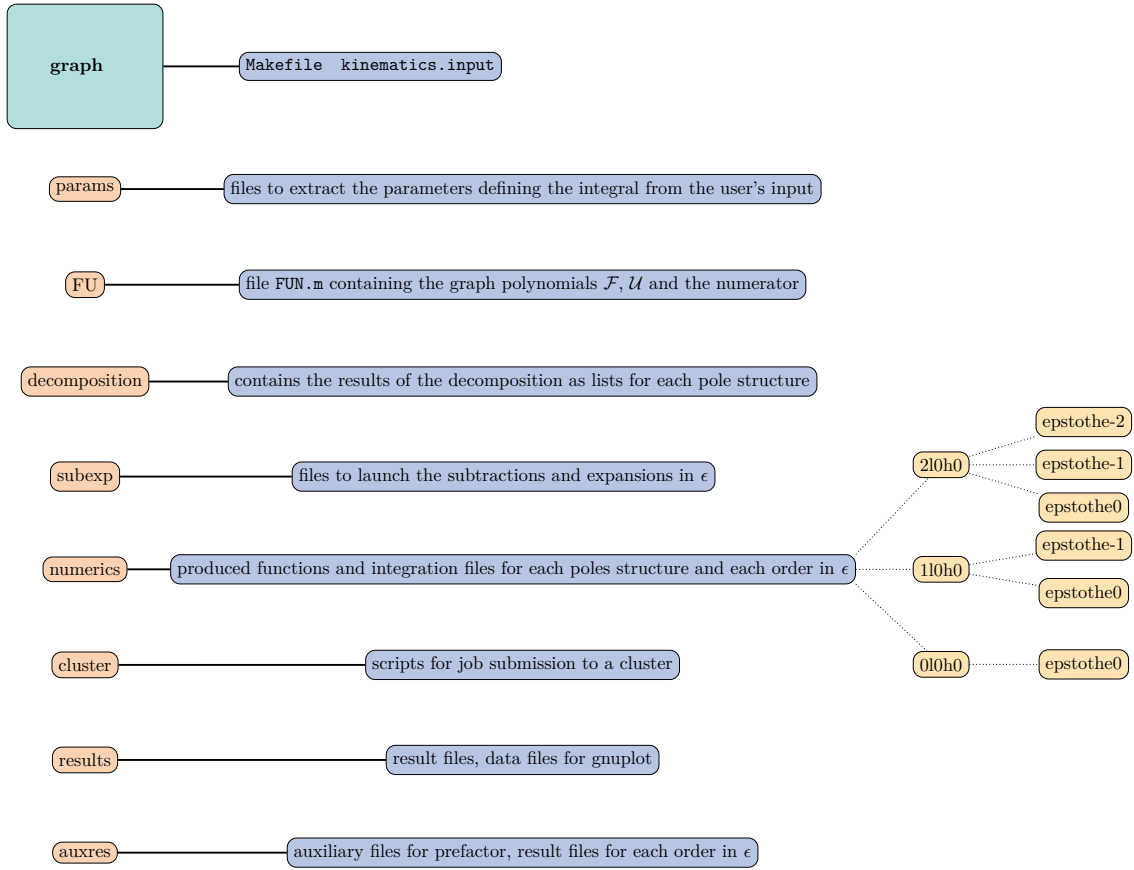}}
\end{picture}
\caption{Output directory structure generated by the algebraic part of \secdec, and example of a pole structure in the {\tt numerics} folder 
containing maximally two logarithmic poles.
	\label{fig:outdirstructure} }
\end{figure}

The output of \secdec is such that the entire output directory generated by the algebraic 
part of \secdec can be transferred as a standalone archive
to another machine or cluster, where the numerical evaluation of all kinematic points can be 
submitted in parallel. Optionally, the user can also evaluate selected pole coefficients individually, see Section~\ref{sec:selectpolecoeff}.
The output directory structure which will be created by \secdec when calculating 
a loop diagram is depicted in
Fig.~\ref{fig:outdirstructure}.

\clearpage

\subsubsection{Decomposition strategies}
\label{sec:strategy}

It is well known that the iterated sector decomposition algorithm 
can run into an infinite recursion if the variables to be rescaled are chosen 
in an inconvenient way. 
As an example, consider the function 
\begin{equation}
f(x_1,x_2,x_3)=x_1^2+x_2^2\, x_3\;,
\end{equation}
and suppose we choose to rescale $x_1$ and $x_3$. The replacement 
$x_1=x_3\,t_1$ in the subsector associated with $\theta(x_3-x_1)$ leads to
$\tilde{f}=x_3\,(x_3\,t_1^2+x_2^2)$. Substituting now $x_2=x_3\,t_2$  
in the corresponding subsector 
remaps to the original 
functional form, so we generate an infinite recursion. 
In this simple example we also see that the 
problem is avoided by choosing to rescale $x_1$ and $x_2$ instead.

The problem can be studied systematically by mapping it to a problem in algebraic geometry, 
which in turn allows to find procedures which are guaranteed to terminate.
This was first noticed in Ref.~\cite{Bogner:2007cr} by mapping it to Hironaka's polyhedra game, 
and then refined in Refs.~\cite{Kaneko:2009qx,Kaneko:2010kj}, see also Ref.~\cite{Smirnov:2008aw}.

In {\sc SecDec}-3.0, we have implemented two additional decomposition strategies, which are based on the 
formalism outlined in Refs.~\cite{Kaneko:2009qx,Kaneko:2010kj} and are therefore guaranteed to stop.

{\sc SecDec}-3.0 uses the program {\sc Normaliz}~\cite{2012arXiv1206.1916B,Normaliz} for the calculation 
of convex polyhedra and their triangulations, needed for the geometric decomposition strategies.

The first method {\tt G1} is an implementation of the original algorithm by Kaneko and Ueda. 
Details can be found in Refs.~\cite{Kaneko:2009qx,Kaneko:2010kj}.

The second decomposition algorithm {\tt G2}, while also based on the ideas of Refs.~\cite{Kaneko:2009qx,Kaneko:2010kj}, 
differs from the original algorithm.
Strategy {\tt G2} uses the Cheng-Wu theorem~\cite{Cheng:1987ga,Smirnov:2006ry} to integrate out the $\delta$-distribution in Eq.~\eqref{eq:scalarloopint}, 
instead of the primary sector decomposition employed in other
decomposition strategies.
The Cheng-Wu theorem states that, instead of the $\delta$-distribution
in Eq.~(\ref{eq:scalarloopint}), it is also possible to use
$\delta(1-\sum_{l\in {\cal S}} x_l)$, where ${\cal S}$ is a
subset of Feynman parameter labels occurring in the integral, 
provided that the integrations for the
remaining Feynman parameters, i.e. those with labels not in ${\cal
  S}$, are performed from zero to infinity. In our implementation we
chose ${\cal S}=\{N\}$, which amounts to setting the parameter $x_N$
equal to one, and integrating over $x_1,\ldots,x_{N-1}$ from zero to
infinity. Even though this breaks possible symmetries among Feynman
parameters,  it turned out to be beneficial with regards to the
overall number of produced functions.

Following this procedure, the Minkowski sum of the Newton polytopes of $\mathcal{U}$, $\mathcal{F}$ 
and the numerator are calculated. The Newton polytope of a polynomial is defined as the 
convex hull of its exponent vectors. Besides the definition as the convex hull of a set of points, a polytope $\Delta$ 
can also be specified as an intersection of half-spaces~\cite{Oda1988}:
\begin{equation}
\Delta = \bigcap_{F} \left\{ \mathbf{m}\in\mathbb{R}^{N-1} \mid \langle \mathbf{m},\mathbf{n}_F\rangle + a_F \geq 0 
\right\}\;.
\end{equation}
Here $\mathbf{n}_F$ denotes the primitive normal vector of the facet labeled by $F$, while $a_F$ is its distance to the origin. 
The index $F$ runs over all facets of the polytope $\Delta$.
{\sc Normaliz} implements efficient algorithms to translate between the two descriptions.

For each sector $j$ a variable transformation
\begin{equation}
x_i=\prod_{F\in S_j} y_F^{\langle \mathbf{e}_i,\mathbf{n}_F \rangle}
\end{equation}
is performed which remaps the integration variables to the unit hypercube. Here the standard basis 
of $\mathbb{R}^{N-1}$ is given by the vectors $\mathbf{e}_i$. 
The sets of facets $S_j$ are chosen in the following way:\\
For each vertex of the polytope, $S_j$ contains the facets incident to it.
If the vertex lies in more than $N-1$ facets, the set is decomposed further using a triangulation. 
The number of generated sectors depends on the way this 
triangulation is performed. \secdec uses the triangulation algorithm of {\sc Normaliz}.
\begin{table}
\centering
\begin{tabular}{|p{3cm}|p{3cm}|p{3cm}|p{3cm}|}
\hline
Diagram & {\tt X} & {\tt G1} & {\tt G2} \\
\hline
\includegraphics[width=3.cm,trim= 0 0 0 -10,clip]{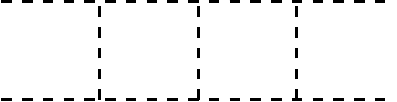} & $282$\newline$1$\,s & $266$\newline$8$\,s & $166$\newline$4$\,s \\
\hline
\includegraphics[width=3.cm,trim= 0 0 0 -10,clip]{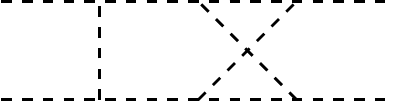} & $368$\newline$1$\,s & $360$\newline$9$\,s & $235$\newline$5$\,s \\
\hline
\includegraphics[width=3.cm,trim= 0 0 0 -10,clip]{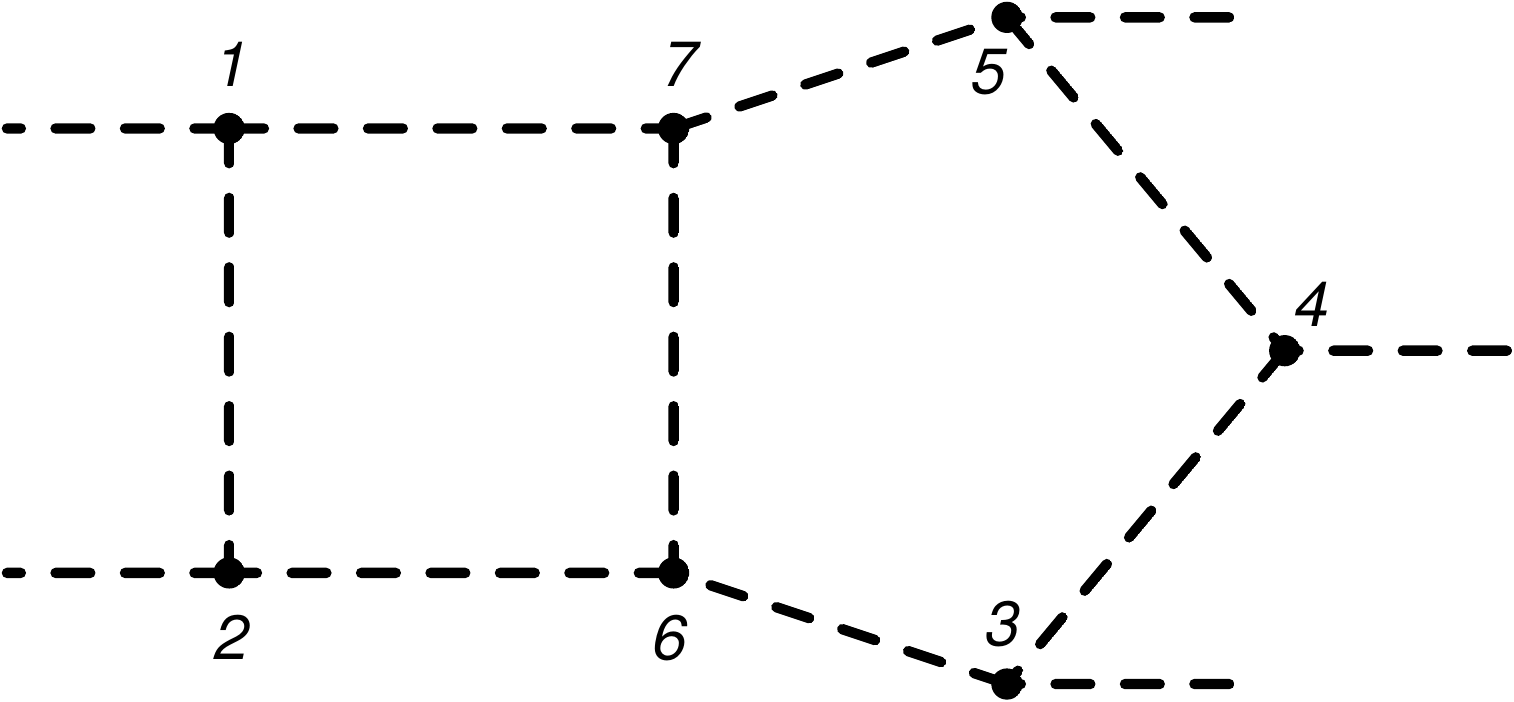} & $548$\newline$3$\,s & $506$\newline$15$\,s & $304$\newline$4$\,s \\
\hline
\includegraphics[width=3.cm,trim= 0 0 0 -10,clip]{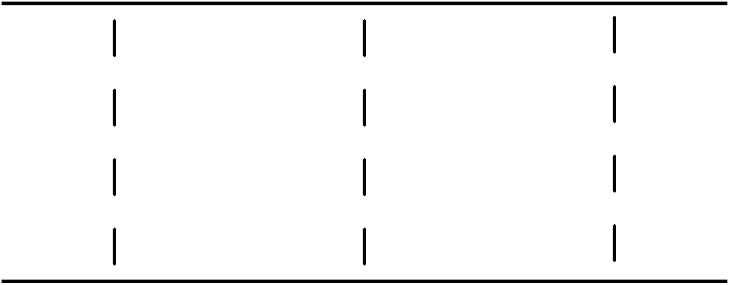} & infinite\newline recursion & $72$\newline$5$\,s & $76$\newline$1$\,s \\
\hline
\includegraphics[width=2.5cm,trim= 0 0 0 -2,clip]{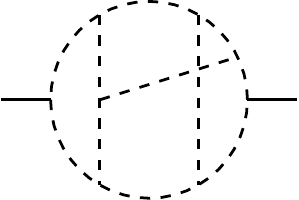} & $27336$\newline$5510$\,s & $32063$\newline$11856$\,s & $27137$\newline$443$\,s  \\
\hline
\end{tabular}
\caption{Number of sectors produced by the implemented decomposition strategies and timings obtained with 
our implementation of the algorithm.
Dashed lines denote massless propagators.\label{tab:sectortable}}
\end{table}

Table \ref{tab:sectortable} compares the number of sectors produced by the decomposition algorithms implemented in {\sc SecDec}-3.0.
The geometric strategy {\tt G2} usually generates the lowest number of sectors.
In cases where the integrals have spurious poles, which cancel in the final result,
strategy {\tt X} can make up for its larger number of sectors by producing less spurious poles. 
\subsubsection{Improvements in speed}
\label{sec:speed}
Towards the goal of using \secdec for the computation of a large number of master integrals
occurring in amplitudes beyond one loop, it is vital to improve on the numerical evaluation times. 
Several steps towards this goal have been undertaken in version 3 of the 
program. 
\subsubsection*{CQuad}
As the integrators included in the {\sc Cuba} 
library~\cite{Hahn:2004fe,Agrawal:2011tm,Hahn:2014fua} are optimized for 
multi-dimensional parameter integrals, it is advantageous to include 
an additional numerical integrator which is dedicated to 
a very fast evaluation of one-dimensional integrals. One of the currently fastest 
and still adaptive integrators is {\sc Cquad}\,\cite{cquad}. 
A version written in the C programming language is included in the 
{\small GSL library}~\cite{Gough:2009:GSL:1538674}. To use it as a 
standalone tool and to make the \secdec package as slim as possible, 
only the necessary GSL files are included in our new release.

\subsubsection*{Compilation times}
The setup of the program is such that the algebraic part needs to be 
performed only once for a given diagram, creating functions where the kinematic invariants
or other parameters  are still in symbolic form.
Their numerical values only need to be specified for the numerical 
integration, without the need to redo any algebraic step 
(except if special kinematics are chosen which would change the singularity structure, 
e.g. massive lines which would lead to IR singularities in the limit $m\to 0$ 
cannot be assigned a zero mass value).

In previous versions of the program, numerical integration files for each kinematic point 
were generated. This is evaded in the new version, allowing to compile 
the resulting functions once and for all, passing the numerical values as arguments. 
This has a substantial effect on the performance, as was noticed in the 
inclusion of the program \secdec into {\tt FeynHiggs}~\cite{Heinemeyer:1998yj,Heinemeyer:1998np,Degrassi:2002fi,Frank:2006yh,Hahn:2009zz,Hahn:2013ria} 
to compute 34 mass configurations of several two-loop two-point integral topologies 
in an automated way, see Ref.~\cite{Borowka:2014wla}. For the latter calculation, 
a preliminary version of the one presented in this paper was used.

\medskip

Apart from these changes, we also altered the type of timings displayed in the 
result files. While in previous versions of the program, the processor time spent 
on only the numerical integration part was measured in clock ticks per second, we now 
switched to wall clock times 
displaying the real time needed for the numerical routine (including 
the presampling when the contour deformation is switched on). 
Even though the real times vary depending on how many other processes a user 
is currently running, we found this information more valuable to the user. 
%
%
\subsubsection{Inverse propagators}
\label{sec:inversepropagators}
As an alternative to the possibility to specify the numerator of a loop integral as a list of 
scalar products of loop and external momenta, propagators with negative (integer) powers are 
supported now, which can be convenient depending on the way the calculation is organized.

The implementation of such inverse propagators relies on the following generalization of the
Feynman parametrization, which accounts for negative propagator powers by calculating 
the derivative instead of the integral of the corresponding parameter~\cite{Smirnov:2008py}.
For a $D$-dimensional $L$-loop integral with $N$ propagators and indices 
$\nu_1,\cdots,\nu_N$, of which only $N_p$, say the first $N_p$, are positive, 
it reads
\begin{align}
  \label{eq:invprops}
  G &= (-1)^{N_{\nu,p}}  \Gamma\left(N_\nu-\frac{LD}{2}\right)
  \int_0^\infty\prod_{i=1}^{N_p} \dd{x_i} \frac{x_i^{\nu_i-1}}{\Gamma({\nu_i})}
  \deltafun\left(1-\sum_{k=1}^{N_p} x_k\right) \nonumber \\
  &\quad \cdot \prod_{j={N_p+1}}^N \frac{\partial^{|\nu_j|}}{\partial x_j^{|\nu_j|}}
  \frac{\mathcal{U}^{N_\nu-\frac{(L+1)D}{2}}}{\mathcal{F}^{N_\nu-\frac{LD}{2}}} \Bigg|_{x_{N_p+1}=\cdots=x_N=0},
\end{align}
where $N_\nu$ denotes the sum of all positive and negative indices, whereas $N_{\nu,p}$ is only the sum
of the positive ones.
The functions $\mathcal{F}$ and $\mathcal{U}$ are defined in the usual way, including, however,
the propagators with negative powers as well.
After performing the derivatives, the result can be written in the form
\begin{align}
  G &= (-1)^{N_{\nu,p}}  \Gamma\left(N_\nu-\frac{LD}{2}\right)
   \int_0^\infty\prod_{i=1}^{N_p}\dd{x_i} \frac{x_i^{\nu_i-1}}{\Gamma({\nu_i})}
  \deltafun\left(1-\sum_{k=1}^{N_p} x_k\right) \nonumber \\
  &\quad \cdot \frac{\mathcal{U}_0^{N_\nu-N_{\nu,n}-\frac{(L+1)D}{2}} \mathcal{N}}
                {\mathcal{F}_0^{N_\nu+N_{\nu,n}-\frac{LD}{2}}},
\end{align}
where $\mathcal{U}_0=\mathcal{U}|_{x_{N_p+1}=\cdots=x_N=0}$, $\mathcal{F}_0=\mathcal{F}|_{x_{N_p+1}=\cdots=x_N=0}$,
$N_{\nu,n}=\sum_{j={N_p+1}}^N|\nu_i|=N_{\nu,p}-N_\nu$, 
and $\mathcal{N}$ is a polynomial in $x_1,\cdots,x_{N_p}$, which is calculable in terms of 
the partial derivatives of $\mathcal{F}$ and $\mathcal{U}$ up to degree $N_{\nu,n}$.

The validity of Eq.~\eqref{eq:invprops} can be seen easily making a detour over to Schwinger parameters.
Depending on the sign of its index, one can introduce a Schwinger parameter $x$ for a propagator 
$P^{-1}$ via~\cite{Smirnov:2006ry}
\begin{align}
  \frac 1{ P^\alpha} = 
  \begin{cases}
    \frac{(-1)^\alpha} {\Gamma(\alpha)} \int_0^\infty \dd{x} x^{\alpha-1} e^{x P} & \alpha >0 \\
    \frac{\partial^{|\alpha|}}{\partial x^{|\alpha|}}e^{x P}\big|_{x=0} & \alpha \leq 0\;.
  \end{cases}
\end{align}
Performing the usual procedure of completing the square in the exponential, shifting and integrating
out the loop momenta, it can be shown that the Schwinger-parametrized form of $G$ 
in terms of $\mathcal{F}$ and $\mathcal{U}$ is given by
\begin{align}
  \label{eq:invprops_Schwinger}
  G &= (-1)^{N_{\nu,p}} 
  \int_0^\infty \prod_{i=1}^{N_p} \dd{x_i} \frac{x_i^{\nu_i-1}}{\Gamma({\nu_i})}
  \prod_{j={N_p+1}}^N \frac{\partial^{|\nu_j|}}{\partial x_j^{|\nu_j|}} 
  \mathcal{U}^{-\frac{D}{2}} \exp\left(-\frac{\mathcal{F}}{\mathcal{U}}\right)\Bigg|_{x_{N_p+1}=\cdots=x_N=0}.
\end{align}
This equation can be transformed to Feynman parameters by inserting
\begin{align}
1=\int_0^\infty\dd{r}\delta\left(r-\sum_{i=1}^{N_p} x_i\right) 
\end{align}
and rescaling $x_i\to r x_i$ for $i=1,\cdots,N$ afterwards~\cite{Smirnov:2006ry}.
Finally one performs the integral over $r$ using the relation
\begin{align}
\int_0^\infty\dd{r} r^{N-1} \exp\left(-r c \right) = \frac{\Gamma\left(N\right)}{c^N}
\end{align}
to complete the proof of Eq.~\eqref{eq:invprops}.

\subsubsection{Linear propagators}
\label{subsec:linprops}

As mentioned already, \secdec can also deal with ``propagators" which do not have a quadratic, 
but only a linear term in the loop momentum. 
They occur for example in heavy quark effective theory or in non-covariant gauges.
An example is given in Section \ref{subsec:examplelinprops}, where we calculate
\begin{equation}
I=\int \frac{d^Dk}{i\pi^\frac{D}{2}}\frac{1}{(k^2+i\,\delta)((k-p_1)^2+i\,\delta) (2k\cdot v+i\,\delta)}
\label{eq:linprop}
\end{equation}
with $v^2\not=0$, $v\cdot p_1=0$. 

Please note that the $+i\,\delta$ term in the linear propagator $(2k\cdot v+i\,\delta)$ 
fixes the analytic continuation prescription. The contour deformation implemented in \secdec 
is constructed under the assumption that each propagator carries a  $+i\,\delta$ term. 

We should emphasize in this context  that great care 
needs to be taken when calculating integrals in the light-cone gauge.
In this gauge, one has an additional linear term $q\cdot n$ in the denominator, 
where $q$ is the loop momentum and 
$n$ is the gauge fixing auxiliary vector with $n^2=0$.
However, the ``naive" $+i\delta$ prescription is problematic
in the light-cone gauge. Therefore the so-called 
{\it Principal Value (PV)} prescription~\cite{Frenkel:1976bia,Curci:1980uw} has been suggested,
\begin{equation}
\frac{1}{q\cdot n}\to \lim_{\delta\to 0}
\left(\frac{1}{q\cdot n+i\,\delta}+\frac{1}{q\cdot n-i\,\delta}\right)=
\lim_{\delta\to 0}\frac{q\cdot n}{(q\cdot n)^2+\delta^2}\;,
\label{eq:pv}
\end{equation}
but it is not compatible with Wick rotation, 
which makes the calculations based on this prescription quite cumbersome.
The only prescription which places the poles in the complex $q_0$ plane 
in the same way as for the  propagators in covariant gauges when integrating over the 
energy component of the loop momentum, and therefore 
preserves the causality behaviour, is the {\it Mandelstam-Leibbrandt (ML)} 
prescription~\cite{Mandelstam:1982cb,Leibbrandt:1983pj,Bassetto:1998uv}:
\begin{equation}
  \frac{1}{q\cdot n}\to \lim_{\delta\to 0^+}\frac{1}{q\cdot n+i\,\delta\,\mathrm{sign}(q\cdot n^*)}=
\lim_{\delta\to 0^+}\frac{q\cdot n^*}{(q\cdot n) (q\cdot n^*)+i\,\delta}\;.
\label{eq:ml}
\end{equation}
However, this prescription introduces the dual vector 
$n^*$ ($(n^*)^2=0, n\cdot n^*\not=0)$ and therefore leads to a result 
containing more invariants than in the presence of only one auxiliary vector $n$.

The above prescriptions are not compatible with the automated contour deformation 
as implemented in \secdec, therefore integrals in the light-cone gauge 
only can be evaluated in the Euclidean region.

\subsubsection{Using Mathematica NIntegrate}

For the numerical integration part, the user of version 3 can choose to either use an integrator from 
the {\sc Cuba} library~\cite{Hahn:2004fe}, or Mathematica's {\tt NIntegrate}. The available Mathematica options to 
the latter command can also be specified directly in {\tt param.input}.
This feature is available in the {\tt loop} setup, while in the {\tt general} setup, the 
Monte Carlo integrator {\sc Bases}~\cite{Kawabata:1995th} is available in addition to {\sc Cuba}.
An example for the usage of {\tt NIntegrate} is given in Section~\ref{subsec:examplenintegrate}.

\subsubsection{Epsilon-dependent dummy functions}

This is a new feature which is available in the {\tt general} branch. 
It allows to define dummy functions which depend on the integration variables, 
but are not explicitly included in the process of iterated sector decomposition. 
The dummy  functions are included as external functions in the numerical integration step, 
where the correct replacement of the integration variables, 
which have been rescaled during the iterated decomposition, 
has been performed automatically.
This can be very useful in cases where large, but finite, expressions should be 
integrated together with a function where the poles need to be factorized. 

In version 2 of the program, such a feature was already present, but it did not 
allow to have $\eps$-dependent expressions to be masked by the dummy functions.
Now the dummy functions can depend on $\eps$ and the program makes sure that 
the various orders in $\eps$ will be combined automatically with the corresponding pole coefficients
from the sector decomposition.

%% file: usage.tex
\subsection{Installation}

The program can be downloaded from 
{\tt http://secdec.hepforge.org}.
Unpacking the tar archive via 
{\tt  tar xzvf SecDec-3.0.tar.gz} will create a directory called {\tt SecDec-3.0}. 
Running  {\tt make} in the {\tt SecDec-3.0} directory will call the {\it install} script, 
which will check whether Mathematica and perl are present and 
compile the numerical integration libraries {\sc Cuba}\,\cite{Hahn:2004fe,Agrawal:2011tm,Hahn:2014fua},  
{\sc Bases}\,\cite{Kawabata:1995th} and {\sc Cquad}\,\cite{cquad}, along with the 
quasi-random sequence generator {\sc Sobol}, which come with the package.
Prerequisites are Mathematica~\cite{Wolfram} version 7 or above, perl 
(installed by default on most Unix/Linux systems), 
a C++03 compliant compiler, and a Fortran compiler if the Fortran part is used. 
Contour deformation is not available in Fortran. 

Please note that the current stable release of Mathematica (v10.0.2) has a bug which causes 
non-interactive sessions, such as those used by \secdec, to hang if
parallel kernels are launched. 
Therefore we have set { \tt nbmathsubkrnls=0 } as the default. 
However, for complicated integrands, using  parallelization for the
algebraic part can lead to substantial improvements in speed. 
In such cases, we recommend using a different version of Mathematica 
and enabling the parallel execution by setting { \tt nbmathsubkrnls}
to a non-zero number in the input parameter file.

The program {\sc Normaliz} 2.10.1~\cite{2012arXiv1206.1916B,Normaliz}\footnote{We should note that version 2.10.1 for our purposes works better than the
currently most recent version  2.12.2.} is needed for the geometric decomposition strategies {\tt G1} and {\tt G2}.
Precompiled executables for different systems can be downloaded from
{\tt http://www.math.uos.de/normaliz/Normaliz2.10.1/}
and have to be moved into the {\tt src/} subdirectory of {\sc SecDec}-3.0. 

The user can check whether the installation was successful with the command 
{\tt make check}, which will run a few test examples and compare the results to  
the pre-calculated result coming with the program package.

\subsection{Usage}
\label{sec:usage}

In the following we refer to three directory structures:
\begin{itemize}
\item { \bf input directory}: the directory in which the user's input files (parameter file, math file and kinematics file) reside,
\item { \bf output directory}: the directory into which {\sc SecDec}-3.0 will write output files
\item { \bf \secdec directory}: the location of {\sc SecDec}-3.0.
\end{itemize}

The {\tt secdec} script is located in the main \secdec directory. It 
is recommended to add the path to the {\tt secdec} script  to the default search paths, so that it can be called from anywhere on 
the user's system.
In the following, we assume that {\tt secdec} was added to the search path, so that it can be called without 
always specifying the path to the script explicitly.

In a first step, the user should create templates for the input files using the command\\
{\tt secdec -prep}.\\
The templates will have the default names {\tt param.input}, {\tt math.m}, {\tt kinem.input}.
The user should edit these files to define the integral to be calculated and the parameters for the numerical
integration.  
If the default names for the input files are kept, the program is called from the input directory with the command\\ 
{\tt \textbf{secdec}}.\\
If the user has renamed the input files, the command is\\
{\tt \textbf{secdec}} {\tt -p <parameter\_file> -m <math\_file> -k <kinematics\_file>}.\\
The output directory will be created by {\sc SecDec}. By default the output directory will be a subdirectory of the
input directory and will have the name of the graph specified in the parameter file.

An additional option, which can be very useful if the user would like to factorize the poles 
from functions which do not have the standard Feynman parameter integral format 
including an overall $\delta(1-\sum x_i)$, is the {\it user-defined} option, which is 
invoked by adding a ``{\tt -u}" to the program call, i.e. \\
{\tt secdec  -u}.\\
This option skips the primary decomposition step, as it assumes that 
there is no $\delta$-constraint between the Feynman parameters.
It requires a certain input format for the functions to be decomposed, 
given in the file {\tt math.m}, which is explained in the 
example in Section~\ref{subsec:exampleuserdefined}. 
Contour deformation is available for these functions, where it is assumed that 
the user defined function which takes the place of the graph polynomial ${\cal F}$
can be integrated on a contour following Feynman's $+i\delta$ prescription 
for the deformation of the contour.
Templates for the input files in the {\it user-defined} setup can be generated using the command \\
{\tt secdec -prep -u}.

Likewise, {\tt secdec -prep -g} will generate templates for the {\tt general} setup.

A commented list of all options can be generated by
{\tt secdec -help}.

\subsection{Description of the input files}

{\bf param.input}\\
The only mandatory fields to be edited by the user in {\tt param.input} are 
{\tt graphname}: a name for the integral to be calculated, 
and {\tt epsord}: the order in epsilon the user wishes to expand the result.
All other parameters take default values if not specified. 
If a template for {\tt param.input} is generated by  {\tt secdec -prep}, 
the default values for the options will be filled in already. 
However, they need not be listed in {\tt param.input}. 
The examples in Section~\ref{sec:examples} contain some short input files where 
only the fields relevant to the calculation of this particular diagram are specified
in the input.
All the input options are described in detail in Appendix \ref{subsec:options}.

\vspace*{3mm}

{\bf math.m}\\
The file {\tt math.m} should contain the definition of the graph to be calculated 
(in Mathematica syntax).
The template generated by  {\tt secdec -prep} contains the definitions for a one-loop box diagram.
All the settings are explained in detail in Appendix \ref{appendix:loop}.

\vspace*{3mm}

{\bf kinem.input}\\
This file is needed for the numerical evaluation to assign numerical values to the 
invariants appearing as symbolic parameters in the functions generated during the algebraic part. 
The symbols for the kinematic invariants have been defined by the user in {\tt math.m}
in the lists {\tt KinematicInvariants} and {\tt Masses}.  
Each line in {\tt kinem.input} should correspond to one kinematic point, where the 
order of the numerical values should match the order of the invariants given in these lists, 
and the masses should always come after the invariants formed by Lorentz vectors. 
Further, each line should start with a label for the numerical point (``pointname"), 
such that the job submission files and result files for different kinematic points belonging to 
the same graph can be distinguished. \\
{\bf Example:}\\
{\tt KinematicInvariants = \{s,t\}}\\
{\tt Masses = \{m1sq,m2sq\}}\\
If the user would like to calculate three numerical points, where 
the Mandelstam invariant {\tt s} can take the values 4,6,8, the value for {\tt t} is fixed to -0.3, 
the (squared) masses are 1.4  and 2.5, 
and the points are just labeled by {\tt p1,p2,p3},
then {\tt kinem.input} should look like\\
{\tt p1 4. -0.3 1.4 2.5}\\
{\tt p2 6. -0.3 1.4 2.5}\\ 
{\tt p3 8. -0.3 1.4 2.5}\\
Assuming a case in which the user already has tables of numbers for the invariants, which do not contain a ``pointname" 
in the first row, the script {\tt addpointname} can be used  
to insert these labels into the data files. The script is located in the folder {\tt loop/src/shell}. 
The loop over all kinematic points listed in {\tt kinem.input} is performed automatically. 
Lines which are commented out by a ``\#'' are ignored. 

\subsection{Run modes of the program}
\label{subsec:runmodes}  

The program consists of an {\bf algebraic part}: factorization of the poles, 
a {\bf numerical part}: integration of the pole coefficients, and a
{\bf collection part}: collection of the numeric results.
If the user only would like to use the algebraic part, 
it is sufficient to issue the command\\
 {\tt secdec -algebraic }.\\
Similarly, to perform just the numerical integrations, the command \\
 {\tt secdec -numerics }\\
 should be used. The command\\
 {\tt secdec -collectresults }\\
  finally collects the results. 
 (Any unique substring of the command flags, such as e.g. {\tt secdec -c[ollectresults]} has the same effect.)

 In single machine mode, all these steps will be done automatically in one go if the user sets the 
 {\tt exeflag} in {\tt param.input} to its maximal value (3), which is also the default. 
 
 In cluster mode ({\tt clusterflag=1}), it is not possible to collect the results automatically, 
 as the information whether all jobs are done is not automatically available. 
 Further, the idea is that the user performs the algebraic step once and for all 
 on a single machine (where Mathematica is available), and then transfers the {\tt graph}
 directory to a cluster, where the numerics can be run. 
 Therefore, in cluster mode, the three commands 
 {\tt secdec -algebraic}, {\tt secdec -numerics}, {\tt secdec -collectresults} have to be issued 
 separately.
 
 The various possibilities for the user to control the different stages of the calculation are shown in Table \ref{tab:exetable}. The optional detailed or basic commands must be run in the order presented. If the user specifies an exe flag and calls \secdec without a basic or detailed command then all tasks with a lower exe flag will also be executed. If the user calls \secdec with a basic or detailed command, for example {\tt secdec -subexpand} then only that task will be performed. Specifically, in this example {\tt makeparams}, {\tt makeFU}, {\tt decompose} and {\tt preparesubexpand} would not be executed.
 

\begin{table}
  \hspace{-7mm}
\begin{tabular}{|c|l|l|l|}
\hline
exe- & command & command & functionality\\
flag& (detailed)  & (basic) &\\
\hline
&&\rdelim\}{16}{50pt}[algebraic]&\\
\ldelim\{{6}{24pt}[$\ge 0$]&   & & $\bullet$ extracts parameters from the graph\\
&   makeparams &&\quad  definition given in {\tt math.m}\\
&   makeFU & & $\bullet$ constructs the graph polynomials ${\cal F}$ and ${\cal U}$\\
&   decompose & & $\bullet$ performs the iterated sector decomposition\\
&   preparesubexpand & & $\bullet$ writes the Mathematica files which are   \\
&&&\quad  needed to run the subtractions and \\
&&&\quad $\eps$-expansions for each pole structure\\
&   & & $\bullet$ performs the subtractions and
$\eps$-expansions\\
\ldelim\{{5}{24pt}[$\ge 1$]&&&\quad for each pole structure and  writes the \\
& subexpand&&\quad functions   {\tt f*.cc} (resp. {\tt f*.m}) to be evaluated \\
&&&\quad   numerically for each pole structure\\
&   preparenumerics & & $\bullet$ writes the files needed to perform the\\
&&&\quad  numerical integration of  each pole structure  \\
&&&\quad  (including the scripts for job submission\\
&&&\quad to a cluster)\\
$\ge 2$&   numerics & numerics & $\bullet$ performs the compilation and runs\\
&&&\quad the executables\\
3&   collectresults & collect & $\bullet$ performs the collection of the results\\
&&&\\
\hline
\end{tabular}
\caption{The different execution stages of \secdec and the possibilities for the user to steer them. The optional detailed or basic commands can be issued with, e.g. {\tt secdec -algebraic}, and must be run in the order presented. \label{tab:exetable}}
\end{table}



\subsection{Evaluation of selected pole coefficients}
\label{sec:selectpolecoeff}

In version 3, the user has the possibility to calculate 
certain pole structures/epsilon orders  individually. 
The pole structures are labeled by a string of the type 
{\tt i}\,{\it l}\,{\tt j}\,{\it h}\,{\tt k}, where {\it l} stands 
for linear poles, {\it h} stands for higher than linear poles, 
while the number {\tt i} of ``usual" logarithmic poles is put at the beginning of the string.
For example, the pole structure {\tt 2l1h0} means ``2 logarithmic poles, 1 linear pole, 0 higher than
linear poles". By ``linear pole" we mean 
that a factorized Feynman parameter has an exponent of the form $x^{-2-b\eps}$.
For more details we refer to Ref.~\cite{Carter:2010hi}.

Each pole structure contains several orders in the $\eps$-expansion, 
ranging from the maximal pole of the given pole structure to 
the expansion order in $\eps$ specified by the user. The coefficients 
of a certain order in $\eps$ are stored in 
subfolders of the pole structure folders, which are labeled {\tt epstothe-2}, {\tt epstothe-1}, etc.
The user can select different pole structures as well as individual $\eps$ orders by commands of the form\\
{\tt secdec [-subexpand/preparenumerics/numerics] [-polestructs=}\\
{\tt <polestructs>] [-epsords=<epsords>]}.\\

The  behaviour of the program is the following:
\begin{itemize}
\item if {\tt polestructs} is not specified, loop over all (contributing) pole structures
\item if {\tt epsords} is not specified, loop over all $\eps$ orders
\item if {\tt polestructs} is specified either as a list \\
{\tt secdec [-subexpand/preparenumerics/numerics] -polestructs=}\\
{\tt 2l0h0,1l0h0},  
or separately \\
{\tt secdec [-<subexpand/preparenumerics/numerics>] }\\
{\tt -polestructs=2l0h0 -polestructs=1l0h0}\\
loop over just these pole structures (for all $\eps$ orders)
\item if {\tt epsords} is specified either as a list or separately, 
loop over just these $\eps$ orders (for all contributing pole structures)
\item if both {\tt polestructs} and {\tt epsords} is specified, loop over just these $\eps$ orders and pole structures.
\end{itemize}

If the {\it togetherflag} is set to one, all pole structures have been combined into 
one folder called  {\tt together}. In this case, the individual $\eps$ orders can still be calculated
separately with the same logic as above, except that there is only one pole structure called ``together".
For example, in order to perform the numerical evaluation only for the $\eps^{-4}$ 
part of all orders in $\eps$ contained in the 
{\tt together} folder, the command would be\\
{\tt secdec -numerics  -epsords=-4}.

%% file: examples.tex

The folders {\tt loop/demos} and {\tt general/demos} contain several examples, where some of them 
are tailored to demonstrate a certain new feature, like the usage of a numerical integrator 
from Mathematica instead of the Cuba library ({\tt NIntegrate}), 
the calculation of tensor integrals ({\tt box\_withnumerator\_2L}), 
integrals with $\eps$-dependent propagator powers ({\tt epsprops\_triangle\_3L}), 
or integrals where strategy X does not stop ({\tt geostrategy\_2L}).

The input files for the demos are kept minimal, so the user can see 
which of the input parameters are the relevant ones, 
while the unspecified parameters  take default values.

The results the user should obtain if the example has run correctly are 
given as files with the extension {\tt .res} in the subdirectory {\tt check} 
of the corresponding example.

The timings we give for the examples below have been obtained on an 
Intel(R) i7-4790K @ 4.00GHz machine. 
We give the overall time, i.e. not only the time for compilation and numerical integration,
but also including the time taken by the algebraic part.

\subsection{One-loop box}
\label{subsec:examplen1Lbox}
\begin{figure}[htb]
\begin{center}
\includegraphics[width=3.8cm]{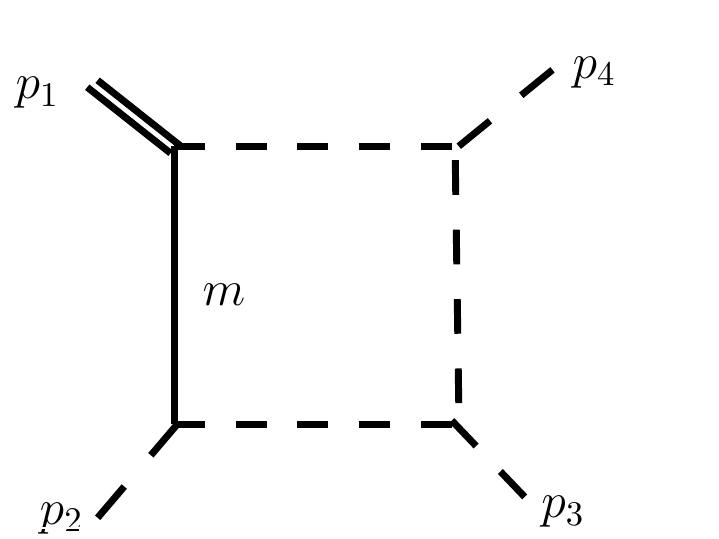}
\end{center}
\caption{The one-loop box example, with $p_1$ being an off-shell leg. Massless
propagators and light-like legs are shown as dashed lines.}
\label{fig:box1L}
\end{figure}

This demo is contained in the folder {\tt 1\_box\_1L}.
It contains a simple example of a one-loop box graph 
with one off-shell leg and one internal mass, see Fig.~\ref{fig:box1L}.
Two kinematic points are calculated, corresponding to the two lines in {\tt kinem.input}.

The overall time to obtain results for the two kinematic points calculated in this example was 41 seconds.

\subsection{Two-loop triangle}

\begin{figure}[htb]
\begin{center}
\includegraphics[width=6.2cm]{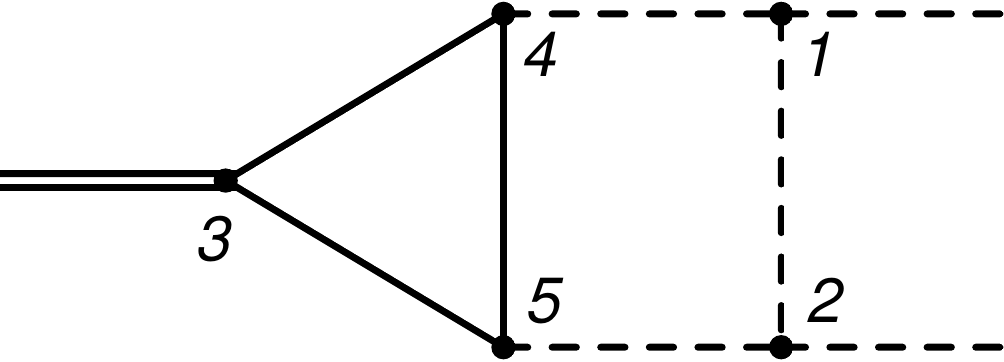}
\end{center}
\caption{The graph {\tt P126}, containing a massive triangle loop.}
\label{fig:P126}
\end{figure}

The example {\tt 2\_triangle\_2L} (denoted by {\tt P126} in version 2) is a two-loop three-point function
containing a massive triangle loop, see Fig.~\ref{fig:P126}.
Analytical results for this diagram can be found e.g. in  Refs.~\cite{Davydychev:2003mv,Bonciani:2003hc}, 
a threshold scan is given in Ref.~\cite{Borowka:2012yc}.

The overall time taken by this example was 144 seconds.

\subsection{Two-loop non-planar box with internal masses}

This example is contained in the folder {\tt 3\_nonplanarbox\_2L}. 

\begin{figure}[ht]
\begin{center}
\includegraphics[width=7.cm]{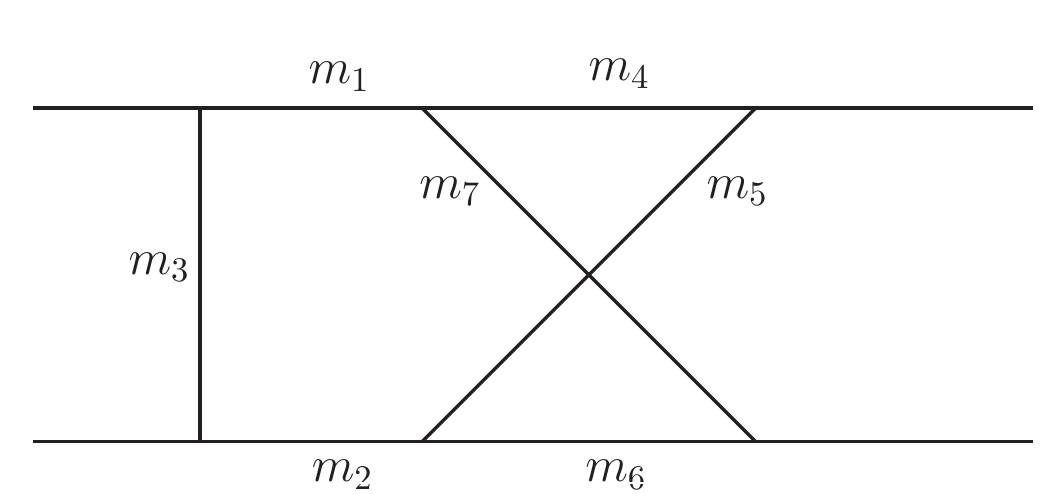}
\end{center}
\caption{Labeling of the masses for the non-planar graph $J^{NP}$.}
\label{fig:JapNP}
\end{figure}

It is a 7-propagator non-planar two-loop box integral where all propagators are massive, 
using $m_1=m_2=m_5=m_6=m$,  $m_3=m_4=m_7=M$, $p_1^2=p_2^2=p_3^2=p_4^2=m^2$. 
The labelling is as shown in Fig.\,\ref{fig:JapNP}. 

Numerical results for this integral have first been calculated in Ref.~\cite{Yuasa:2011ff} using a
method based on extrapolation in the $i\delta$ parameter.
We give results for $m=50, M=90, s_{23}=-10^4$. A scan over the invariant $s_{12}$ can be 
found in Ref.~\cite{Borowka:2012yc}.

The overall time taken by this example was 186 seconds.

\subsection{Two-loop planar box with loop momenta in the numerator}

\begin{figure}[ht!]
\centering
\includegraphics[width=9.cm]{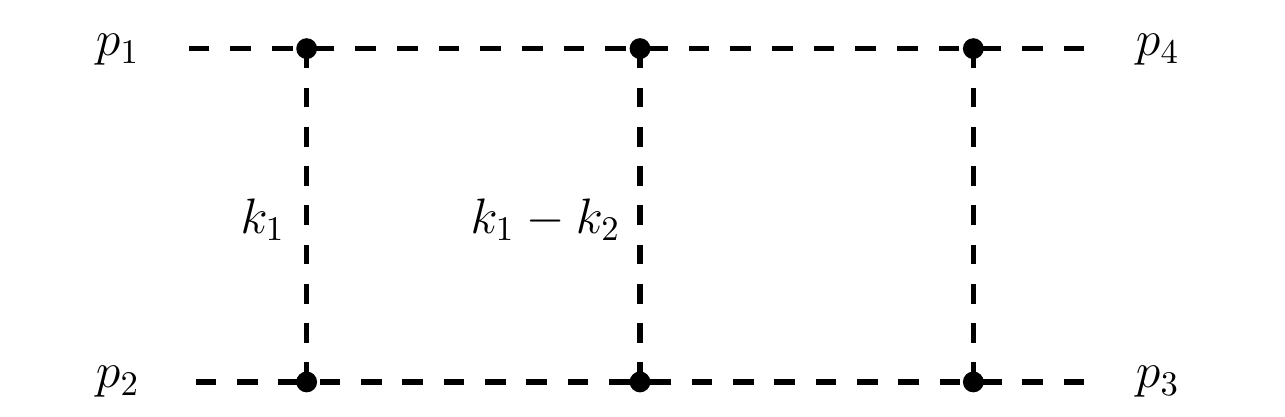}
\caption{Momentum labelling for the two-loop planar box. 
All propagators and external legs are assumed to be massless.} 
\label{fig:boxwithnum}
\end{figure}
In this example, which can be found in the folder {\tt 4\_box\_withnumerator\_2L},
we consider a planar massless two-loop box diagram with an additional
propagator that is raised to a negative power, i.e. is in the numerator.


Using the momentum labelling shown in Fig.~\ref{fig:boxwithnum}, we choose 
\begin{align}
(k_1+p_3)^2 = k_1^2 + 2 k_1\cdot p_3
\end{align}
as the inverse propagator.
There are two ways to compute the integral with this numerator:
\begin{itemize}
\item Add $(k_1+p_3)^2$ to the propagator list and specify its index as $-1$ in the power list.
This method is demonstrated in the default input files in the example folder and can be run
by just typing {\tt secdec}.
\item Calculate two integrals, with $k_1^2$ and $2 k_1\cdot p_3$, respectively, as numerators.
In the first integral, the numerator can be cancelled against a propagator.
This pinched graph can be calculated by simply removing this propagator from {\tt proplist}, or by setting
its index to $0$ in the {\tt powerlist}.
For the second integral, we explicitly specify the numerator in the mathfile {\tt math\_tensor.m}.

These integrals can be run by specifying the corresponding input files:\\
{\tt secdec -m math\_pinched.m -p param\_pinched.input}\\
{\tt secdec -m math\_tensor.m -p param\_tensor.input}
\end{itemize}

We emphasize that the two methods to calculate integrals with numerators cannot be combined, 
i.e. when an index smaller than zero is specified, the {\tt numerator} given in {\tt math.m}
must not include any loop momenta.

The overall time taken by the example was 14 minutes for the graph with inverse 
propagator, 
i.e. the default input files {\tt param.input, math.m}, below 3 minutes for 
the graph with specified numerator and 19 seconds for 
the pinched graph. 

\subsection{Two-loop pentagon}

This example is contained in the folder {\tt 5\_pentagon\_2L}. 
\begin{figure}[ht!]
\centering
\includegraphics[width=0.5\textwidth]{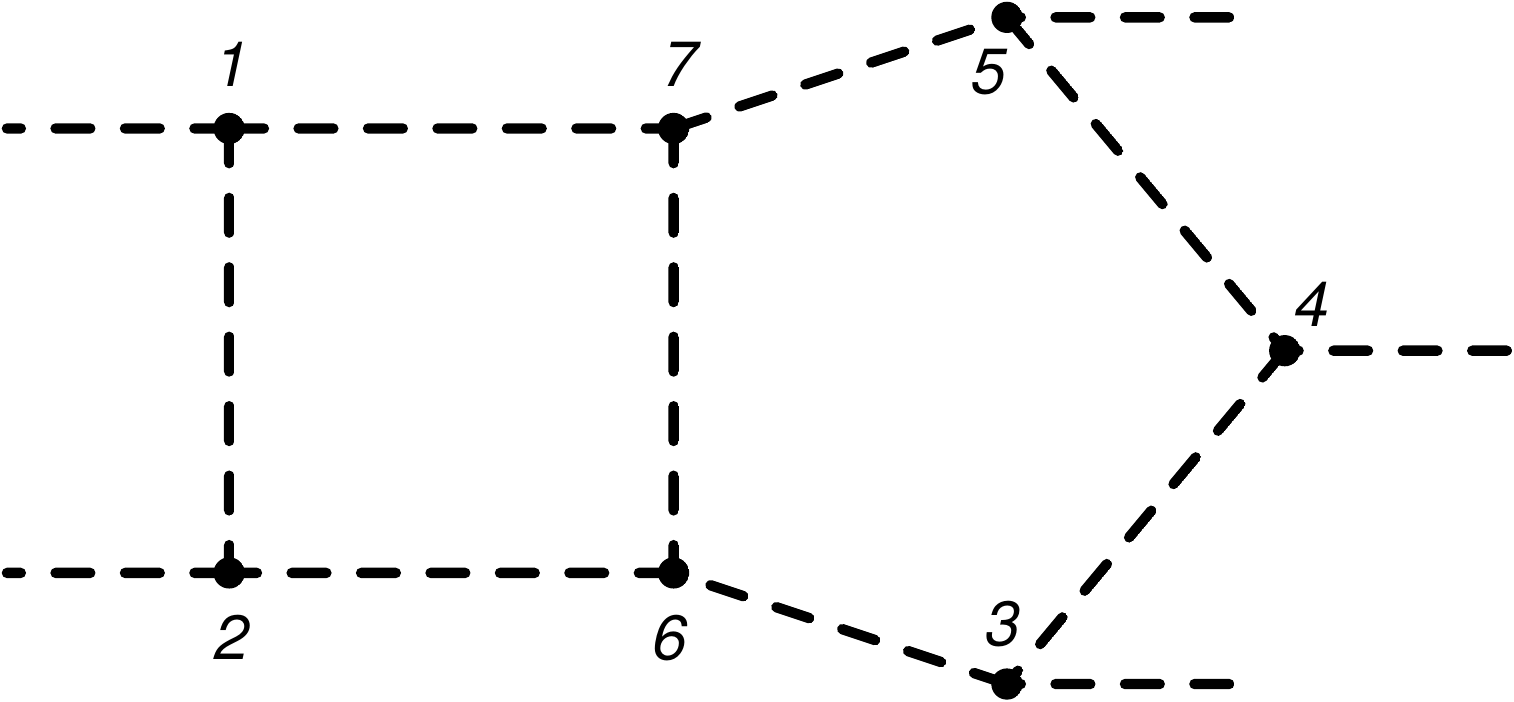}
\caption{The numbers label the vertices of the massless two-loop five-point integral.} 
\label{fig:pentagondiag}
\end{figure}
The calculation of the diagram in Fig.~\ref{fig:pentagondiag} demonstrates the 
applicability of \secdec to examples with more than four external legs. Assuming 
the external momentum $p_i$ to be situated at vertex $i$, 
and choosing the invariants formed by two adjacent legs as a basis, 
the following scalar products need to be defined

{\tt ScalarProductRules = \{\\
  SP[p1,p1]$\to$0, SP[p2,p2]$\to$0, SP[p3,p3]$\to$0, SP[p4,p4]$\to$0, \\
  SP[p5,p5]$\to$0, SP[p1,p2]$\to$s12/2, SP[p1,p3]$\to$(s45-s12-s23)/2, \\
  SP[p1,p4]$\to$(s23-s51-s45)/2, SP[p1,p5]$\to$s51/2,\\
  SP[p2,p3]$\to$s23/2, SP[p2,p4]$\to$(-s23-s34+s51)/2,\\
  SP[p2,p5]$\to$(s34-s12-s51)/2, SP[p3,p4]$\to$s34/2, \\
  SP[p3,p5]$\to$(s12-s34-s45)/2, SP[p4,p5]$\to$s45/2
 \};}

which are formed from relations among the kinematic invariants $s_{ij}=(p_i+p_j)^2$.

The overall time taken for this example was 9 minutes and 43 seconds.

\subsection{Two-loop with geometric decomposition strategy}

\begin{figure}[ht!]
\begin{center}
\includegraphics[width=4.5cm]{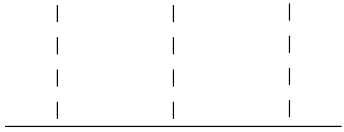}
\end{center}
\caption{The two-loop box example with four massive on-shell legs. Dashed lines denote massless propagators.\label{fig:qed}}
\end{figure}

This example is contained in the folder {\tt 6\_geostrategy\_2L}.
It calculates the two-loop box diagram shown in Fig.~\ref{fig:qed}
using the strategy {\tt G2}. With strategy {\tt X} the decomposition runs into an infinite recursion.
Numerical results for this diagram in the Euclidean region have first been given in Ref.~\cite{Binoth:2003ak}.
Analytical results can be found in Ref.~\cite{Heinrich:2004iq}.
Here we expand the result up to order $\eps^2$.
 
The overall time taken for this example was 41 seconds.

Please note that {\sc Normaliz}~\cite{2012arXiv1206.1916B,Normaliz} must be installed to run this example.

\subsection{Three-loop triangle}

\begin{figure}[ht!]
\begin{center}
\includegraphics[width=4.cm]{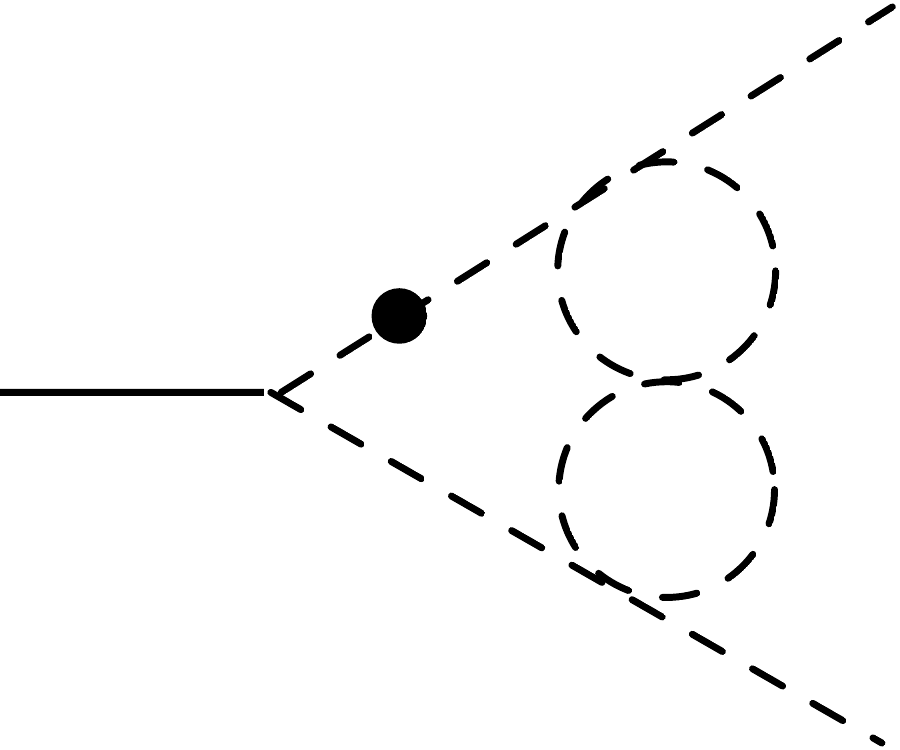}
\end{center}
\caption{The three-loop vertex diagram $A_{6,1}$ with the dotted propagator raised to the power $1+\eps$.\label{fig:a61}}
\end{figure}

This example is contained in the folder {\tt 7\_epsprops\_triangle\_3L}. 
Apart from being a 3-loop example, it also demonstrates the usage of $\eps$-dependent propagator powers.
It calculates the diagram shown in Fig.~\ref{fig:a61}, where one of the propagators is raised to the power
$1+\eps$.
We expand the result up to order $\eps^2$.

The analytical result for this diagram with general propagator powers is given in Ref.~\cite{Gehrmann:2006wg} 
and is also given in the file {\tt A61analytic.m} to allow comparisons between 
analytical and numerical results for arbitrary propagator powers.

The overall time taken for this example was 13 seconds.

\subsection{Linear propagators}
\label{subsec:examplelinprops}

This example is contained in the folder {\tt 8\_linearprop\_2L}. 
It contains the integral 
\begin{equation}
I=\int \frac{d^Dk}{i\pi^\frac{D}{2}}\frac{1}{(k^2+i\,\delta)((k-p_1)^2+i\,\delta) (2k\cdot v+i\,\delta)}
\end{equation}
with $v^2\not=0$, $v\cdot p_1=0$. 

Please note that in the presence of linear propagators, {\tt proplist} must be given 
in the form of explicit momentum flow, i.e. constructing the graph polynomials ${\cal F},{\cal
U}$
from labeled vertices is not possible.

The overall time taken for this example was 30 seconds.

\subsection{Using NIntegrate instead of Cuba} 
\label{subsec:examplenintegrate}

This example is contained in the folder {\tt 9\_NIntegrate\_box\_1L}. 
It is identical to the one-loop box described in Section~\ref{subsec:examplen1Lbox},
except that the {\tt NIntegrate} function from {\tt Mathematica} is used for 
the numerical integration.

The overall time taken for this example was 40 seconds.

\subsection{User defined integrands} 
\label{subsec:exampleuserdefined}

This example is contained in the folder {\tt 10\_userdefined\_triangle\_1L}. 
\begin{figure}[ht!]
\centering
\includegraphics[width=0.3\textwidth]{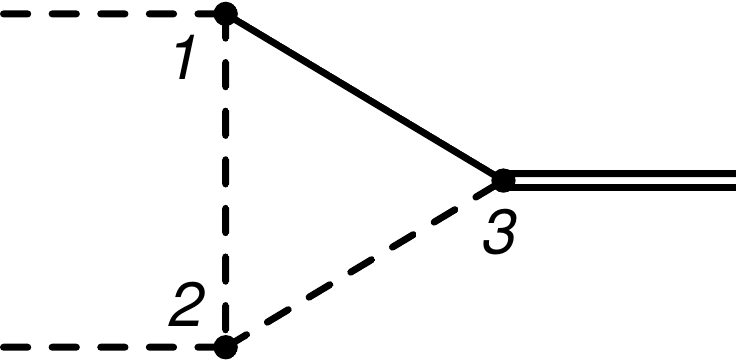}
\caption{The numbers label the vertices of the one-loop three-point integral. 
Dashed lines denote massless internal or external particles, 
full lines denote massive ones.} 
\label{fig:1Ltriangle}
\end{figure}
In this, a one-loop three-point function with one massive propagator and one 
off-shell leg, compare 
Fig.~\ref{fig:1Ltriangle}, is computed using the user-defined setup.

The user-defined setup supports functions of the type 
\begin{align}
G_{\text{user}}= \int\limits_{0}^{1} 
\,\prod\limits_{j=1}^{n}\text{d}z_j\,\,z_j^{\nu_j}\,
{\cal N}\; {\cal U}^{\text{expou}}\; {\cal F}^{\text{expof}} \text{ ,}
\label{eq:userfuncdef}
\end{align}
where $n$ is the number of integration parameters 
$z_j$ to be integrated from zero to one and ${\cal N}$ is a numerator function which  
cannot contain singularities. ${\cal U}$ and ${\cal F}$ may contain singularities and, 
if needed, sector decomposition can be applied to factorize these. Those parts of a 
user function which need an analytical continuation {\`a} la loop integral, should be 
written to the ${\cal F}$ part, as the integration contour is deformed based on the 
properties of this function. It should be noted that, comparing with 
Eq.~(\ref{eq:scalarloopint}), Eq.~(\ref{eq:userfuncdef}) does not contain a 
$\delta$-distribution. 

In the standard Feynman loop integral case, the $\delta$-distribution is integrated 
out during the primary sector decomposition. Performing the primary sector 
decomposition of the one-loop triangle diagram, see Fig.~\ref{fig:1Ltriangle}, by 
hand, the resulting functions read 
\begin{subequations}
\label{subeq:userints}
\begin{align}
G_{1,\text{user}}=&\int\limits_{0}^{1}\text{d}z_1 \text{d}z_2 z_1^{-1-\eps} 
(1+z_1+z_2)^{-1+2\eps}(-s z_2 + m^2(1+z_1+z_2))^{-1-\eps}\;,\\
G_{2,\text{user}}=&\int\limits_{0}^{1}\text{d}z_1 \text{d}z_2 z_2^{-1-\eps} 
(1+z_1+z_2)^{-1+2\eps}(-s + m^2(1+z_1+z_2))^{-1-\eps} \;,\\
G_{3,\text{user}}=&\int\limits_{0}^{1}\text{d}z_1 \text{d}z_2  
(1+z_1+z_2)^{-1+2\eps}(-s z_2 + m^2(1+z_1+z_2))^{-1-\eps}\text{ .}
\end{align}
\end{subequations}
It must be noted that the standard loop prefactor 
$\frac{(-1)^{N_{\nu}}}{\prod_{j=1}^{N}\Gamma(\nu_j)}\Gamma(N_{\nu}-LD/2)$ in Eq.~(\ref{eq:scalarloopint}) is 
not included in the setup for the userdefined functions. In order to get the same result as in 
the Feynman loop integral case, one therefore has to compute
\begin{align}
\nonumber G_{\text{loop}}=&-\Gamma(1+\eps) (G_{1,\text{user}}+G_{2,\text{user}}+G_{3,\text{user}}) \text{ .}
\end{align}
In \secdec, the functions of Eqs.~(\ref{subeq:userints}) are defined in the functionlist
~\\
{\tt functionlist = \{\\
{\small \{1, \{-1-eps,0\}, \\
\{\{(U[z]/.z[1]$\to$1)/.z[3]$\to$t[1], expou, A\}, \\
\{-s*z[2] + msq*(1+z[1]+z[2]), expof, A\}\},
Num\}, \\
\{2, \{0,-1-eps\}, \\
\{\{(U[z]/.z[2]$\to$1)/.z[3]$\to$t[2], expou, A\}, 
\{-s + msq*(1+z[1]+z[2]), \\
expof, A\}\}, 
Num\}, \\
\{3, \{0,0\}, \\
\{\{U[z]/.z[3]$\to$1, expou, A\}, 
\{F[z]/.z[3]$\to$1, expof, B\}\}, 
Num\}\}; }}\\
~\\
where the functions {\tt F[z]} and {\tt U[z]} corresponding to the two graph polynomials 
of the Feynman one-loop triangle integral, their exponents {\tt expou, expof} and the 
numerator {\tt Num} of the integral are separately written to the {\tt math.m} file 
using Mathematica syntax
~\\
{\tt Num = 1;} \\
{\tt expou = -1+2*eps;} \\
{\tt expof = -1-eps;}\\
~\\
{\tt U[z\_]:= z[1] + z[2] + z[3]} \\
{\tt F[z\_]:= z[3]*(-s*z[2] + msq*(z[1] + z[2] + z[3]))} .

The user functions are labeled by their first entry in the {\it functionlist}. If two 
functions share the same exponents {\tt expou} and {\tt expof}, and are labeled 
with the same number, they are summarized and then decomposed as if they belonged 
to one sector. 

The second entry of each user function in the {\it functionlist} is a list of exponents. 
Here, the exponents of each integration parameter should be given. The first 
function $G_{1,\text{user}}$ of this triangle example contains a factor $z_1^{-1-\eps}$ 
but no factor of $z_2$. This is translated into the statement {\tt \{-1-eps,0\}}. 

The third entry of each user function in the {\it functionlist} contains a list of 
information on the functions corresponding to the first and second graph polynomial after primary 
sector decomposition in our example. This is the function itself, its exponent and the 
flag {\tt A} or {\tt B}. While {\tt A} denotes that no further decomposition is needed, {\tt B} leads to 
an iterated sector decomposition of the function. 

While the input files {\tt param.input} and {\tt kinem.input} are the same for the Feynman 
loop integral and the userdefined setup, the {\tt math.m} input file 
has to contain the above mentioned functionlist definition. Unlike the loop setup, 
definitions for {\tt momlist, proplist, numerator, powerlist,} 
{\tt ExternalMomenta, externallegs, ScalarProductRules} need not be included if these are not 
necessary for the definition of the integrand in {\it functionlist}.

To run this example, go to the demo directory {\tt 10\_userdefined\_triangle\_1L} 
and issue the command 
{\tt secdec -u}, the {\tt-u} activates the userdefined setup.


The overall time taken for this example was 7 seconds.

\subsection{Dummy functions}
This example demonstrates the usage of $\eps$-dependent dummy functions in the calculation of general parametric integrals.
It is located in the folder {\tt general/demos}. The integrand of this example reads
\begin{align}
 f= (z_1+z_2)^{-2-2\eps} \,z_3^{-1-4\eps}\, \textrm{dum}_1^{1+\eps}(z_1,z_2,z_3,z_4;\eps)\, \textrm{dum}_2^{2-6\eps}(z_2,z_4)\,\textrm{cut}(z_3),
\end{align}
with
\begin{subequations}
\begin{align}
  &\textrm{dum}_1(a,b,c,d;\eps) = a^2+b^3+c^4+d^5+4 a b c d + 2 -a^2b^3c^4d^5 + 5 a \eps,\\
  &\textrm{dum}_2(a,b) = a^2 + b^2 + \beta^2+4 a b - \sqrt{a b \beta} +3 a^2 b^2, \quad \beta=0.5\\
  &\textrm{cut}(z_3) = \Theta(z_3-C), \quad C=0.1.
\end{align}
\end{subequations}
Note that the dummy functions must not lead to additional singularities.
Before {\tt secdec} can be launched to calculate the integral 
  $\int_0^1 \text{d}z_1\text{d}z_2\text{d}z_3\text{d}z_4\, f,$
Fortran code for the calculation of the dummy functions has to be generated. This can be done by specifying the names of the dummy functions, as well as their highest power in $\eps$, in the parameter file and running\\
{\tt perl createdummyfortran.pl -d demos/ -p dummyfunction.input}\\
from the {\tt general} folder. This will translate the Mathematica files containing the definitions of the dummy functions to Fortran code. However, the file {\tt cut.f} has to be edited afterwards to account for the $\Theta$-function.
These modifications can be found in the file {\tt cut\_edited.f} and should be copied to the file {\tt cut.f}.

The integration is launched with\\
{\tt secdec -g -p dummyfunction.input -m dummyfunction.m}\\
and will take into account the $\eps$-dependence of the dummy functions. When doing the algebraic and numeric part of \secdec separately, it is possible to modify the dummy functions after the algebraic part as long as their $\eps$-dependence does not exceed the maximum power specified in the parameter file. Further details can be found in a readme file located in the {\tt general/demos} directory.
\label{subsec:exampledummy}

The overall time taken for this example was 3 seconds.

%% file: appendix.tex
%

\secdec{} has various setups:
one for standard loop integrals, termed `loop setup' in the following, 
and one for general parametric functions, termed `general setup', 
as reflected by the two directories {\tt loop} and {\tt general}.
While the parametric functions treated in the {\tt general} folder are not accessible to contour
deformation, the latter is still available for integrals which go beyond standard loop
integrals. Such non-standard integrals can be defined by the user, 
and therefore this setup is referred to as `user-defined setup', 
which is an option within the 
{\tt loop} directory.

The program is called by invoking the script {\tt secdec}, located in the main \secdec directory. 
We recommend to add the path to the {\tt secdec} script  to the user's default search paths, 
so that it can be called from anywhere on 
the system.
In the following, we assume that {\tt secdec} was added to the search path, so that it can be called without 
always specifying the path to the script explicitly.

\subsection{Overview of usage in the loop setup}
\label{appendix:loop}

\subsubsection{Basic usage}
\begin{itemize}
\item Create templates for the input files with the command\\
{\tt secdec -prep}\\
This command generates the  files {\tt param.input}, 
{\tt kinem.input} and {\tt math.m}.
\item Input the graph name and desired order in $\eps$ ({\tt epsord}) into {\tt param.input}, 
define the loop integral in {\tt math.m} and specify one or more kinematic points in 
{\tt kinem.input}. Please note that the order of the values for the kinematic invariants
given in {\tt kinem.input} must match the order of the invariants given in {\tt math.m},
i.e. the entries of the lists {\tt KinematicInvariants}, {\tt Masses} (masses always listed last).
\item To run \secdec: simply issue the command \\
{\tt secdec} \\
If the input files have been renamed, the command is \\
{\tt secdec -p <myparam.input>  -m <mymath.m> -k <mykinem.input>}.
\item Switch to the output directory which has been created. It carries the 
{\it graph name} specified in {\tt param.input}. 
The results are in the {\tt results} folder, 
containing a file {\tt graph\_pointname.res}  for each point specified in the 
kinematics file, and  {\tt plotfile<i>.gpdat} files for each $\eps$-order $i$,
where the results for all kinematic points have been appended.
\end{itemize}

\subsubsection{Intermediate usage}

The different stages of \secdec can also be run separately.
Creating and editing the input files as before, one can run the algebraic part only by \\
{\tt secdec -algebraic}.\\
This allows, for example, to get a first idea of the pole structure generated during the decomposition.
One can also examine the ${\cal F}$ and ${\cal U}$ functions in the folder 
named {\tt FU} in the output directory.

The numerical part is run by the command\\
{\tt secdec -numerics}.\\
The results are collected by \\
{\tt secdec -collectresults}.

Options: 
\begin{itemize}
\item to add a few more kinematic points, for example in the threshold region, 
one can edit the {\tt kinem.input} file in the input directory. 
The user should delete the old (already calculated) points from the {\tt kinem.input} file and 
add the new points with new point names. 
\item One can re-run the numerics in cluster mode by uploading the output directory to the head node and 
running the submission scripts in the {\tt cluster} folder. 
(Note: please examine the submission scripts before submitting them to your cluster, 
they may need to be adapted to the particular cluster setup).
To collect the results, one should copy the output directory back to the original machine 
(for example by using {\tt rsync}) and then run: {\tt secdec -collectresults}. 
\end{itemize}

\subsubsection{Expert usage}
The various stages of \secdec can be subdivided even more and can be run separately 
by issuing the detailed commands listed in Table~\ref{tab:exetable}.
For example, to only extract the graph parameters and construct the 
graph polynomials ${\cal F}$ and ${\cal U}$, one has to issue the commands\\
{\tt secdec -makeparams} and then \\
{\tt secdec -makeFU}.\\

The user can select pole structures and $\epsilon$ orders to be computed.
For example, the command:\\
{\tt secdec -numerics -polestructs=2l0h0,1l0h0 -epsords=-2,-1}\\
would compute only the contribution of the {\tt 2l0h0} and {\tt 1l0h0} pole structures to
the $\epsilon^{-2}$ and $\epsilon^{-1}$ poles.

For all possible commands we refer to Table~\ref{tab:exetable}.
Please note: if  an {\tt exe flag} is specified and  {\tt secdec} is called without a basic or detailed command 
then all tasks with a lower exe flag will also be executed. 
If the user calls {\tt secdec} with a basic or detailed command, 
for example {\tt secdec -subexpand} then only that task will be performed.

\subsubsection{Description of the input files}
The description below is for loop diagrams; the input files in the subdirectory {\tt general}
to compute more general parametric functions did not change between versions 2 and 3,
and therefore we refer to descriptions in Refs.~\cite{Borowka:2012yc,Borowka:2014aaa} for the {\tt general} branch.

\begin{itemize}
\item {\tt param.input}: (text file)\\
The mandatory parameters the user needs to specify are
\begin{itemize}
\item {\tt graphname}: a name for the graph to be calculated
\item {\tt epsord}: the desired expansion order in $\eps$.
\end{itemize}
All other values take defaults if not specified.
A detailed description of all options is given is Section~\ref{subsec:options}.
\item {\tt math.m}: (Mathematica syntax) \\
This file contains the definition of the graph to be calculated. \\
{\tt momlist}: list of loop momenta\\
{\tt proplist}: list of propagators, either in momentum flow representation 
or as a list containing the propagator mass and the labels of the vertices the propagator is connecting. 
Please note that the label of the vertex which contains the external momentum $p_i$ should be $i$. 
For vertices involving only internal lines, the labelling is arbitrary. For more details we refer to 
Ref.~\cite{Borowka:2012yc}.\\
{\tt numerator}: there are two possiblities to define a numerator:\\
(a) give a list of loop momenta contracted with either
          external momenta or loop momenta. Each Lorentz contraction should be 
	  denoted by ``*", while each contracted factor should form an element of a list. 
	  E.g., for  $ 2\,k_1\cdot p_1\,k_2\cdot p_2$, the syntax is
          {\tt numerator={2,k1*p1,k2*p2}}. \\
(b) define an additional propagator and specify a negative index in {\tt powerlist} (see below).\\	  
The default is numerator={1} (scalar integral).  \\
{\tt powerlist}: list of porpagator powers, also called ``indices" in the literature.
Can also take zero or negative integer values.\\
{\tt Dim}: the dimension of the loop momenta. The default is Dim=$4-2\eps$.\\
{\tt prefactor}: the prefactor specified here will be factored out of the numerical result.
This means that the numerical result will be divided by the prefactor given here.
Please note that a factor $\frac{(-1)^{N_{\nu}}}{\prod_{j=1}^{N}\Gamma(\nu_j)}\Gamma(N_{\nu}-LD/2)$,
  which comes from the Feynman parametrisation, 
  will be included by default in the numerical result, according to the integral definition in
  Eq.~(\ref{eq:integraldef}).\\
{\tt ExternalMomenta}: list of external momenta occurring in the graph definition
(only necessary if the graph definition contains the momenta explicitly, i.e. in momentum flow
representation).
If the length of this list is different from the number of external legs, 
give also the number of external legs as {\tt externallegs=\ldots}.\\
{\tt KinematicInvariants}: list of symbols for kinematic invariants 
(formed from Lorentz vectors) occurring in the diagram.
The symbols can be chosen by the user.\\
{\tt Masses}: list of symbols for the masses. In the case of massive on-shell lines, 
i.e. $p^2=m^2$, where $m$ is a propagator mass as well as the mass of an external leg, 
the  way to proceed is to define {\tt SP[p,p]}$\to$ {\tt m2} in {\tt ScalarProductRules}, such that no extra symbol needs
to be specified for $p^2$. \\
Note that the list of masses should always contain the propagator masses used in {\tt proplist}.\\
{\tt ScalarProductRules}: list of replacement rules for the kinematic invariants formed by external momenta.\\
{\tt splitlist}: 
allows to specify  those Feynman parameters which may have an endpoint singularity 
at $z_i=1$, such that the integral will  be split at 1/2 and the singularity at 
 one will be remapped to a singularity at zero.

\item {\tt kinem.input}: (text file) \\
Should contain numerical values for the kinematic invariants, {\bf in the same order} 
as the symbols for the kinematic invariants given in the fields \\{\tt KinematicInvariants}
and {\tt Masses} 
in the Mathematica input file (called {\tt math.m} here).
The numerical values for the masses should always be listed {\it after} the invariants formed from Lorentz vectors.

If kinemloop.input contains several lines, each line will be 
evaluated as a new kinematic point.
In order to be able to distinguish the runs/results for the different kinematic points, 
each array of numerical values should have a label prepended which defines the ``pointname".
Example: if {\tt mathloop.m} contains {\tt KinematicInvariants = \{s,t,p1sq\}} and 
{\tt Masses=\{m1sq\}},
the corresponding kinematics input file where {\tt m1sq} takes the value 1 for point {\tt p1} 
and 5 for point {\tt p2} should look like (the values for s,t,p1sq are e.g. 500, -88, 100)\\
{\tt p1 500 -88 100 1}\\
{\tt p2 500 -88 100 5}

\end{itemize}

\subsubsection{Detailed description of all options in {\tt param.input}}
\label{subsec:options}

It should be emphasized again that the only mandatory fields are {\tt graph} and {\tt epsord}, 
all other parameters take default values if not specified, which are given in brackets after the
keyword in the description below.

\begin{description}
\item[graph]
The name of the diagram or parametric function to be computed is specified here. 
The graph name can contain underscores and numbers, but should not contain commas.
\item[epsord]
The order to which the Laurent series in $\eps$ should be expanded, 
starting from $\eps^{-maxpole}$. 
The value of {\it epsord=0} will calculate the pole coefficients and the finite part. 
Note that epsord can be negative if only the pole coefficients up to a certain order should be computed.
\item[outputdir ()]
specifies the name of the directory where the produced files will be written to, the {\bf absolute path} should be given.
If left empty, a subdirectory of the input directory with the name of the graph will be created.
\item[contourdef (False)]
The contour deformation can be switched on or off by choosing 
{\it contourdef=True/False} (lower case letters are also possible). 
For multi-scale problems, respectively diagrams with non-Euclidean kinematics, 
set {\it contourdef=True}.
\item[lambda (1.0)]
$\lambda$ is a parameter controlling the contour deformation.
The program takes the $\lambda$ value given by the user in the input
file as a starting point. The program then performs checks and optimizations 
to find an `optimal' value for $\lambda$.
The user should pick an initial value which is 
rather large, as the program will decrease $\lambda$ appropriately, 
while it cannot increase  $\lambda$.
Values of $\lambda$ between 1 and 3 are usually a good choice.
\item[strategy (X)]
Choice of the decomposition strategy. The default is {\tt X}, which is the same strategy as 
in previous versions, which is usually the most efficient one, but is not guaranteed to stop.
If the decomposition does not seem to stop, strategies {\tt G1} or {\tt G2} should be chosen, 
which are based on a geometrical algorithm as described in section \ref{sec:strategy}, 
and which cannot run into an infinite recursion. Within strategy {\tt G2}, no primary sector
decomposition is done. Therefore it is obviously not possible to specify only selected 
primary sectors to be calculated when using this strategy.
\item[exeflag (3)]
The {\it exeflag} can be used to execute the program only up to a certain stage.
There are three basic stages: i) the algebraic part, ii) the numerical part and iii) 
collecting the results. The algebraic part can be further split into a part doing only the iterated 
sector decomposition (e.g. to get an idea about the pole structure), 
and a part performing the subtractions and the expansion in epsilon.
The values of the {\it exeflag}  correspond to the following stages:
\begin{itemize}
\item 0: The parameters like the number of loops, propagators etc are extracted from the 
user's Mathematica input file; the graph polynomials ${\cal F}$ and ${\cal U}$ (and the numerator in the case of tensor integrals) are constructed; the 
iterated sector decomposition is done; the scripts  {\tt subandexpand*.m} in the {\it graph} subdirectory
for the subtractions and epsilon expansions are created, but not run. 
\item 1: The  subtractions and expansions in $\eps$ are performed 
and the resulting functions for the pole coefficients are written to C++ or Mathematica files; 
all the other files needed for the numerical integration are created as well.
\item 2: Compilation and numerical integrations are performed.
\item 3: The results are collected.
\end{itemize}
All exeflags imply that the steps corresponding to lower exeflags will automatically be performed as well.
However, there is also the possibility to skip previous steps by calling \secdec with a basic or detailed command, as explained in Section \ref{subsec:runmodes}. For example, to run only the numerical part the call is {\tt secdec -numerics}.
This command will {\it not} restart the whole algebraic part, but just compile the functions and run
the executables. An error will be thrown if the functions have not been produced beforehand. 
Table~\ref{tab:exetable} gives a schematic overview of the various options to control the program flow.

Please note that if the {\tt clusterflag} is switched on, it is assumed that the user will produce the functions locally
and then compile and run them on a cluster. Therefore, in cluster mode, the program will perform the algebraic part 
and produce the submission scripts, and then stop, independent of the value of the {\tt exeflag}, as the user should 
control the job submission ({\tt secdec -numerics}) and the collection of the results 
({\tt secdec -collect}) in cluster mode.

\end{description}

\subsection*{Advanced usage}
\begin{description}

\item[togetherflag (0)]
This flag defines whether to integrate subsets\footnote{The subsets are 
naturally formed by the fact that functions contributing to a certain $\eps$-order 
can descend from different pole structures.}  of functions 
contributing to a certain 
$\eps$-order separately ({\it togetherflag=0}), 
or to sum all functions for a certain  order in $\eps$ prior to integration 
({\it togetherflag=1}). The latter will avoid large 
cancellations between results for functions descending from different pole structures and thus give a more 
realistic error.
However, {\it togetherflag=1} is not recommended for cases  
where the individual functions are already very complicated.
\item[grouping (2000)]
It can be beneficial to first sum a few 
functions before integrating them. 
Choosing a value for the grouping which is nonzero  
defines an upper limit (in kilobytes)  for a file containing the sum of a number of functions. 
The number of kbytes is set by {\it grouping=\#kbytes}. 
Setting {\it grouping=0}, all functions {\tt f*.cc} resp. {\tt f*.m} are integrated 
separately. In practice, grouping=0 has proven to lead to faster 
convergence and more accurate results in most cases.
However, for integrals where large cancellations among 
the different functions occur, the grouping value should be chosen $\neq0$. 
The log files {\tt *results*.log} in the results directory contain 
the results from the individual sub-sector integrations. These files are useful
to spot cancellations between the individual functions and to adjust the settings accordingly.
\item[IBPflag (0)]
Set {\it IBPflag=0} if the integration by parts option should not be used and {\it IBPflag=1} if it should be used. 
{\it IBPflag=2} is designed to use IBP relations when it is deemed efficient to do so. 
Using the integrations by parts method takes more time in the subtraction and 
expansion step and generally 
results in more functions for numerical integration. Its usefulness is mainly for cases
where (spurious) linear poles of the type $x^{-2-b\eps}$ are found 
in the decomposition, as it reduces the power of $x$ in the denominator.
\item[infinitesectors ()]
This field should be empty if the default strategy {\tt X} or the strategies {\tt G1} or {\tt G2} are applied.
It offers the possibility to use a different `heuristic' strategy~\cite{Binoth:2003ak} 
for certain primary sectors
if they seem to suffer from infinite recursion. 
The primary sectors given in the list (comma separated) will then be decomposed 
with this heuristic strategy. It can avoid infinite recursion in cases where strategy {\tt X} fails, 
but is not guaranteed to stop. 
For example, {\it infinitesectors=2,3} 
results in the application of this heuristic 
strategy  to primary sectors 2 and 3. In examples with massive propagators, one should put the labels 
belonging to the massive propagators  into the list.
\item[primarysectors ()]
This field allows to calculate selected primary sectors only.
If left blank, {\it primarysectors} defaults to all, i.e. {\it primarysectors=1,...,N} will 
be assumed, where $N$ is the number of  
propagators. This option is useful if a diagram has symmetries such that some primary 
sectors yield the same result. It cannot be used in combination with strategy {\tt G2}.
\item[multiplicities ()]
Specify the {\it multiplicities} of the primary sectors listed above. List the {\it multiplicities} in same order as 
the corresponding sectors above. If left blank, a default multiplicity of $1$ is set for 
each primary sector.
\item[rescale (0)]
If there are large differences in magnitude in the kinematic invariants occurring in a diagram,  
it can be beneficial to rescale all invariants by the largest one in order 
to reach faster convergence during the numerical integration. The 
rescaling  can be switched on with {\it rescale=1} (default is {\it rescale=0}).
{\it Please note:} If switched on, it is not  possible to set explicit values (numbers) for any 
non-zero invariant in the {\it ScalarProductRules=} conditions in the Mathematica 
 file {\tt math.m}.
\item[nbmathsubkrnls (0)] 
Maximal number of Mathematica kernels to be used by Mathematica 
(the iterated decomposition and subtractions are not  parallelized by default).

\item[smalldefs (0)]
 {\it smalldefs=1} minimizes the deformation of the contour. It can be useful 
for example in the presence of oscillatory integrands.
\item[largedefs (0)]
If the integrand is expected to  have (integrable) endpoint singularities 
at $x_j=0$ or 1, {\it largedefs=1}, can help to 
have a sufficiently large deformation close to the endpoints.
The default is {\it largedefs=0}. \\
Note that setting both flags {\it largedefs} and {\it smalldefs} to zero 
is perfectly possible, as the flags operate on different parts of the deformation internally. 
For more details we refer to Ref.~\cite{Borowka:2012yc}.
\item[optlamevals (4000)]
The number of pre-samples to determine the optimal contour 
deformation parameter $\lambda$ can be chosen by assigning a number to 
{\it optlamevals}.
\end{description}


\subsection*{Parameters related to the numerical integration and external libraries}
\begin{description}
\item[compiler (gcc)]
Choice of the C-compiler.
\item[CCargs (-O)]
Compiler options for the C-compiler to compile the numerical integration files.
\item[sobolpath ()]
The path to {\tt sobol} can be specified here, if different from the default
{\tt [path\_to\_secdec]/src/sobol}. The {\tt sobol} quasi-random number generator
is only used if {\it contourdef=True}. 
\item[cquadpath ()] 
The path to {\tt cquad} can be specified here, if different from the default
{\tt [path\_to\_secdec]/src/cquad}. 
Note that the program will use this integrator automatically if an integral or 
a pole contribution
is found to depend on only 1 Feynman parameter, irrespective of what has been chosen 
as integrator below.
\item[integrator (3)]
The program for the numerical integration can be chosen here. 
Vegas ({\it integrator=1}), Suave ({\it integrator=2}), 
Divonne ({\it integrator=3}) and Cuhre ({\it integrator=4}) 
are part of the \textsc{Cuba} library.
To choose a numerical integrator included in Mathematica, 
{\it integrator=5} can be chosen. 
\end{description}

\subsection*{NIntegrate parameters}
\begin{description}
\item[NIntegrateOptions ({\tt AccuracyGoal->3})] Options for the Mathematica NIntegrate command, if {\it integrator=5} is
chosen.  Example:\\
NIntegrateOptions=AccuracyGoal $\to$ 2,WorkingPrecision$\to$12,\\
Method$\to$``AdaptiveMonteCarlo"\\
Please note that when using NIntegrate, it is not possible to obtain an error estimate.
\end{description}

\subsection*{Cuba parameters}
Below we describe only a few \textsc{Cuba} parameters, for more information the user 
is referred to the {\sc Cuba} documentation,
Refs.~\cite{Hahn:2004fe,Hahn:2014fua}.
As we have set Divonne to be the default integrator, we give some more
details for Divonne settings below.

\begin{description}
\item[cubapath ()]
The path to the \textsc{Cuba} library can be specified here, if different from the 
default  {\tt [path\_to\_secdec]/src/Cuba-4.1}. 
\item[cubacores (1/0)] The maximal number of cores  Cuba is allowed to use. 
In cluster mode, the default is 1, in single machine mode the default is zero, which means that Cuba 
will use all available idle cores.
\item[maxeval (10000000)]
The maximal number of evaluations to be used by the numerical integrator.
For this field and the fields {\tt mineval, epsrel, epsabs} below, 
a value can be specified for each order in $\eps$,
separated by commas and starting with the leading pole.
If only one value is given, it will be used for all pole coefficients.
If the list is shorter than the number of orders in $\eps$, the last value of the list will be
repeated as often as necessary. 
\item[mineval (0)]
The number of evaluations which should at least be done before the numerical integrator returns a result. 
\item[epsrel ({\tt 1.e-2})]
The desired relative accuracy for the numerical evaluation. 
\item[epsabs ({\tt 1.e-6})]
The desired absolute accuracy for the numerical evaluation. 
These values are particularly important when either the real or the imaginary 
part of an integral is close to zero. 
\item[cubaflags (2)]
Sets the \textsc{Cuba} verbosity flags. The default is 2, which means that the \textsc{Cuba} 
input parameters and other useful 
information, e.g. about numerical convergence, are written to the log file of the 
numerical integration.
\item[seed (0)]
The seed used to generate random numbers for the numerical integration with Cuba.
The default is {\it seed=0}: Cuba will use the Sobol (quasi-) random number generator

\medskip
{\it Divonne specific parameters}
\medskip

\item[key1 (1500)]
Determines the number and type of sampling to be used for the partitioning phase in Divonne. 
With a positive {\tt key1} different from 7,9,11,13, a Korobov quasi-random sample of {\tt key1} points is used.
\item[key2 (1)]
Determines the number and type of sampling to be used for the final integration phase in Divonne. 
With   $n_2=|key2|<40$, the number of sampling points is
$n_2n_{need}$, where $n_{need}$ is the number   of points needed to reach the prescribed accuracy as estimated 
by Divonne from the partitioning phase.
\item[key3 (1)]
Sets the strategy for the refinement phase in Divonne. 
Setting {\tt key3=1} (default), each subregion is split once more.
\item[maxpass (4)]
Number of iterations that are performed before the partition is
accepted as final, to pass on to the main integration phase. 
\item[border ({\tt 1.e-6})]
The points in the interval $[0,\rm{border}]$ and $[1-\rm{border},1]$ are not included in the integration 
but are extrapolated from samples in the interior. This can be useful 
if the integrand is known to be peaked close to endpoints of some integration variables.
\item[maxchisq ({\tt 1.})]
The maximally allowed $\chi^2$ a particular subregion is allowed to
have in the final integration phase. Regions which fail this test and
whose sample averages differ by more than {\tt mindeviation} move on
to the refinement phase.
\item[mindeviation ({\tt 0.15})]
Determines if a region which failed the $\chi^2$ test is treated
further (see {\tt maxchisq} above).
\end{description}


\subsection*{Parameters related to the cluster mode}
\begin{description}

\item[clusterflag (0)]
Determines how jobs are submitted. Setting {\it clusterflag=0} (default) the jobs will run on
a single machine, with {\it clusterflag=1} the jobs will run on a cluster 
(the corresponding files to submit jobs to a cluster will be created, see below).
\item[batchsystem (0)]
Chooses a format for the scripts steering the submission to a cluster.
If {\it batchsystem} is set to 0, the setup is 
for the PBS (portable batch system). If the flag is set to 1,
a user-defined setup is activated. Currently this is the submission via {\tt condor}, but it 
can be easily adapted to other batch systems by editing the templates in {\tt loop/src/numerics/} and 
{\tt loop/perlsrc/makejob.pm}. 
\item[clusteroptscompile ()] 
In cluster mode: command line arguments passed verbatim to the job submission script for compilation jobs on a cluster
\item[clusteroptsrun ()]
In cluster mode: command line arguments passed verbatim to the job submission script for 
numerical integration jobs on a cluster.

\end{description}
\subsection*{Parameters related to plotting}
\begin{description}

\item[xplot (1)]
This option can be used to control the format of the data files where the results 
for a range of kinematic points are listed. The variable {\tt xplot} 
denotes a position in the list of invariants. The corresponding invariant then will be the one 
which will be plotted on the x-axis.
Example: the invariants are s,t,u,m1sq,m2sq. 
If a scan over m1sq has been performed, such that m1sq should be plotted on the x-axis, 
then xplot=4 would tell \secdec to write the values for m1sq into the first column 
of the {\tt *.gpdat} file.
The {\tt *.gpdat} files produced by the program have the form \\
{\tt [invariant chosen by xplot]  real\_result  real\_error imag\_result\\  imag\_error  timing}.\\
For 3D plots: if xplot is a list of length L, the first L columns of {\tt *.gpdat} will 
correspond to the 
values of the invariants singled out by the xplot labels (e.g. xplot=1,2 would produce data
files for a 3D plot in s,t).

\end{description}

\subsubsection{User-defined setup}
This setup allows to define
functions  in the Mathematica input file \\({\tt math\_userdefined.m})
which are not standard loop integrals. It is invoked by the option {\bf -u} when calling {\tt secdec}.
While the {\tt param.input} file has the same form as for standard loop integrals,
the file {\tt math[userdefined].m} should contain the specification of the user-defined 
functions. Most fields are the same as in {\tt math[loop].m}. 
However, instead of the definition of a graph, the user can define 
a list of functions to be decomposed. The detailed format is specified 
in the example described in Section \ref{subsec:exampleuserdefined}, also 
contained in the {\tt demos} folder of the program.
\subsubsection{Looping over ranges of parameters}
\label{subsec:multinumerics}
In order to do the numerical integration for a whole set of numerical points, 
the {\tt multinumerics} script which was present in version 2 has become obsolete
in the loop setup. For this purpose, the user only needs to specify 
numerical values for the kinematic invariants in {\tt kinem.input}, where each line 
defines a new kinematic point.

\subsection{General setup}
The structure of the directory {\tt general} has changed only slightly in version 3 of the program.
The command to launch \secdec is {\tt secdec -g} similar to the {\tt loop} case.
Templates for the input files can be generated by {\tt secdec~-prep~-g}.

The command {\tt secdec -g  -p <param.input>  -m <math.m>} will calculate the integral
using the parameters specified in {\tt param.input}. To evaluate several points one can add 
the additional option {\tt  -k <kinem.input>} to the above command.

The commands \\{\tt secdec -g   -algebraic} 
and\\ {\tt secdec -g   -numerics} 
will also work in the {\tt general} setup.

\begin{itemize}
\item {\tt param.input}: (text file)\\
In this file the user needs to specify a name for the functions to be evaluated,
the desired order in $\eps$, the parameters for the numerical integration,  
and he can specify further options.
The format is similar to {\tt param.input} in the loop integral case,
except that by default all parameters occurring in the function are specified in the file {\tt param.input}
and no {\tt kinem.input} file is needed. However, latter can be used to facilitate parameter scans, following
the syntax of the file {\tt multiparamfile} of \secdec version 2.
\item {\tt math.m}: (Mathematica syntax) \\
Contains the definition of the integrand and further options.
\end{itemize}
%

%% file: secdec3_main.bbl
\providecommand{\href}[2]{#2}\begingroup\raggedright\endgroup